\documentclass[twocolumn,showpacs,aps,prb,superscriptaddress,amsfonts]{revtex4}

\pdfoutput=1

\usepackage{epsfig}
\usepackage{amsmath}
\usepackage{amssymb}
\usepackage{bm}
\usepackage{color}

\newcommand{\beq}{\begin{equation}}
\newcommand{\eeq}{\end{equation}}
\newcommand{\beqarray}{\begin{eqnarray}}
\newcommand{\eeqarray}{\end{eqnarray}}

\newcommand{\eq}[1]{(\ref{#1})} 
\newcommand{\fig}[1]{figure~\ref{#1}} 
\newcommand{\Sec}[1]{section~\ref{#1}} 
\newcommand{\Ref}[1]{Ref.~\onlinecite{#1}} 

\begin{document}

\allowdisplaybreaks

\title{Magnetic order in orbital models of the iron pnictides}  
\author{P. M. R. Brydon}
\email{brydon@theory.phy.tu-dresden.de}
\affiliation{Institut f\"{u}r Theoretische Physik, Technische Universit\"{a}t  
  Dresden, 01062 Dresden, Germany }
\author{Maria Daghofer}
\email{m.daghofer@ifw-dresden.de}
\affiliation{IFW Dresden, P.O. Box 270116, 01171 Dresden, Germany}
\author{Carsten Timm}
\email{carsten.timm@tu-dresden.de}
\affiliation{Institut f\"{u}r Theoretische Physik, Technische Universit\"{a}t  
  Dresden, 01062 Dresden, Germany }

\date{\today}

\begin{abstract}
We examine the appearance of the experimentally-observed stripe
spin-density-wave magnetic order in five different
orbital models of the iron pnictide parent compounds. A restricted
mean-field ansatz is used to determine the magnetic phase diagram of each
model. Using the random phase approximation, we then check this phase
diagram by  evaluating the static spin
susceptibility in the paramagnetic state close
to the mean-field phase boundaries. The momenta for
which the susceptibility is peaked indicate in an unbiased way the actual
ordering vector of the nearby mean-field state. The dominant orbitally
resolved 
contributions to the spin susceptibility are also examined to
determine the origin of the magnetic instability. We find that the
observed stripe magnetic order is possible in four of the models, but
it is extremely sensitive to the degree of the nesting between the electron and
hole Fermi pockets. In the more realistic five-orbital models, this order
competes with a strong-coupling incommensurate state which
appears to be 
controlled by details of the electronic structure below the Fermi energy. We
conclude by discussing the implications of our work for the origin of the
magnetic order in the pnictides.    
\end{abstract}


\maketitle

\section{Introduction}

The origins of high-$T_c$ superconductivity is one of the most important
problems in contemporary condensed matter physics. There are now two main
material classes in which such a state occurs: the cuprates, and the
recently-discovered iron pnictides.~\cite{Kamihara2008,Rotter2008} The latter
have therefore been the subject of an intensive research effort over the past
few years, spurred by the hope
to gain insight into the mechanisms of the
unconventional superconductivity in both families. 

The proximity 
of antiferromagnetic (AFM) and superconducting states in the phase diagram of
the pnictides and 
the cuprates hints that exchange of
virtual spin fluctuations is responsible for the pairing in both
systems.~\cite{1111coupling,122coupling,Huang2008,Lee2006} Although this already
provides strong 
motivation to study the magnetism of the pnictides, the unusual magnetic
properties are of interest in their own right.~\cite{Lumsden2010,Johnston2010}
It is now well established that the undoped cuprates are AFM Mott
insulators.~\cite{Lee2006} In contrast, there is convincing 
evidence that the 
iron pnictide parent compounds $R$FeAsO and $A$Fe$_2$As$_2$ ($R$ and $A$ are
rare-earth and alkaline-earth elements, respectively) display a
\emph{metallic}
spin-density-wave (SDW)
state:~\cite{Lumsden2010,Johnston2010} the relatively low value of
the ordered moment~\cite{delaCruz2008,Huang2008} and the appearance of
incommensurate magnetism with doping;~\cite{Pratt2011} 
the presence of several electron and hole Fermi surface above 
$T_{N}$, and their significant reconstruction in the AFM
state, as revealed by angle
resolved photoemission spectroscopy (ARPES) and quantum oscillation
experiments;~\cite{magneto,Yi2009b,Shimojima2010} the compounds continue to
show metallic transport properties below
$T_{N}$;~\cite{McGuire2008,Liu2008,Dong2008} and  spectroscopic
measurements suggest  
intermediate correlation
strengths.~\cite{Drechsler2009,WLYang2009,chalcogenides}

The microscopic description of the AFM state of the parent compounds in the
cuprates and the pnictides will therefore be qualitatively
different. While the $t$-$J$ model gives a good account of the
former,~\cite{Lee2006} there 
is currently no generally-accepted model of the latter. By treating the
Fe-As planes as a lattice of localized
moments, several 
authors~\cite{Yildirim2008,Uhrig2009,Krueger2009,Schmidt2010} have obtained
good agreement with neutron-scattering data.~\cite{Zhao2009} Although
such a picture may be appropriate for the more strongly correlated iron
chalcogenides,~\cite{Johnston2010,Ma2009} it is  
difficult to reconcile with the experimental evidence for a
moderately correlated metallic system in the 1111 and 122 pnictides.
A more promising starting point for the pnictides is the prediction of
\emph{ab  
  initio} calculations that the nesting between the electron
and hole Fermi pockets derived from the Fe 3$d$ orbitals is responsible for
the SDW.~\cite{nesting,Zhang2010} 
Theoretical work based on this scenario can be divided into two schools of
thought: those which argue that only the nesting is
important,~\cite{Korshunov2008,Chubukov2008,Han2008,Vorontsov2009,Brydon2009a,Cvetkovic2009,Brydon2009b,Thomale2009,Knolle2010a,Eremin2010,Knolle2010b,Fernandes2010} 
and those which also attempt to account for the complicated orbital structure
of the Fermi surfaces.~\cite{Kuroki2008,Lorenzana2008,Daghofer2008,Raghu2008,Lee2008,Yanagi2008,Ikeda2008,Kaneshita2009,Graser2009,Kariyado2009,Kubo2009,Arita2009,Ran2009,SLYu2009,Yu2009,Moreo2009,Kuroki2009,Kaneshita2010,Bascones2010,Daghofer2010a,Ikeda2010,Daghofer2010b,Long2010,Thomale2010,Luo2010,Yang2011,Zhou2010,Kubo2010,Schickling2010}

The first approach can be dubbed the ``excitonic'' theory, as it is based upon
the excitonic instability of a semimetal proposed in the
1960s.~\cite{Excitonic} Within this scenario, the Coulomb attraction between
electrons and holes about the nested electron and hole Fermi pockets, 
respectively, causes the condensation of interband electron-hole pairs
(excitons). The condensed system is a density wave state with ordering vector
${\bf Q}$ equal to the nesting vector of the Fermi surfaces, with secondary
interactions stabilizing a SDW.~\cite{Chubukov2008,Buker1981} Using 
suitably chosen phenomenological bands and effective interactions, the
excitonic theory naturally explains
the reconstruction of the Fermi surface and the continued metalicity below
$T_N$,~\cite{Cvetkovic2009,Eremin2010,Knolle2010b} it can reproduce
the results of neutron-scattering experiments,~\cite{Brydon2009b,Knolle2010a}
and it appears to capture key aspects of the competition between
superconductivity and the AFM
state.~\cite{Chubukov2008,Vorontsov2009,Cvetkovic2009,Thomale2009,Fernandes2010}   

Despite the success of the excitonic approach, it is unsatisfying as a
microscopic model of the iron 
pnictides. Specifically,
it ignores the complicated orbital structure of the different Fermi
surfaces,~\cite{Shimojima2010,Boeri2008,Eschrig2009,ARPESZhang2009} 
and the use of effective interactions obscures the roles of the various
intra-ion interaction terms.~\cite{Chubukov2008,Cvetkovic2009}
Although the basic physics driving the AFM is expected to remain the same, it
is desirable to include the orbital physics for a number of reasons: the
variability in superconducting properties of the doped compounds suggests that
small details of the electronic structure could play a significant role in the
physics,~\cite{Ishida2009} the likely
presence of moderate correlations implies that some local physics should be
included in theoretical models,~\cite{Si2009,Aichhorn2009} while the
real-space structure of the orbital wavefunctions, and the observed
orbital reconstruction of the Fermi surface below $T_N$,~\cite{Shimojima2010} 
indicates a strong coupling of the orbital and magnetic degrees of freedom.

A large number of different orbital tight-binding models have therefore been
proposed involving between two or five of the Fe 3$d$
orbitals,~\cite{Daghofer2008,Raghu2008,Graser2009,Arita2009,Ran2009,Yu2009,Kuroki2009,Daghofer2010a,Ikeda2010,Luo2010,Zhou2010,Calderon2009a,Calderon2009b,Brydon2011}
or also including the As 4$p$ orbitals~\cite{Yanagi2008,Manousakis2008} and
even orbitals from outside the FeAs planes.~\cite{Papaconstantopoulos2010}
Although the existence of superconductivity in these models has been
extensively studied,~\cite{Kuroki2008,Lee2008,Yanagi2008,Ikeda2008,Graser2009,Arita2009,SLYu2009,Kuroki2009,Ikeda2010,Thomale2010,Luo2010,Yang2011} 
the magnetic behaviour remains rather poorly understood. The most
popular approach to the appearance of AFM order in these models is
to examine the ground state using a standard mean-field ansatz that allows
for at most two-site magnetic unit
cells;~\cite{Kaneshita2009,Kubo2009,Ran2009,Yu2009,Bascones2010,Daghofer2010a,Daghofer2010b,Luo2010}
very recently, this has been generalized to a Gutzwiller mean-field
theory.~\cite{Zhou2010,Schickling2010}
As all of the orbital models have a rather complicated electronic structure,
however, it is by no means certain that such phases have the lowest free 
energy, i.e. states with
larger unit cells or incommensurate ordering vectors cannot be \emph{a priori}
excluded. Indeed,
in several cases more advanced techniques have been used to 
study the magnetic properties with mixed results: whereas in some models the
findings are consistent with the usual mean-field
ansatz,~\cite{Kariyado2009,Kaneshita2010,Ikeda2010,Luo2010,Kubo2010}
striking divergences have been found in others.~\cite{Luo2010}
The existence and robustness of the experimentally-observed stripe
AFM order in these models at parent 
compound filling therefore remains an important open problem, as it
is questionable to use them 
to examine the superconductivity of the doped system if
they do not give the correct magnetic behaviour of the parent compound.

It is hence desirable to examine the magnetic properties of these 
orbital models in an \emph{unbiased} way. This can be
achieved by examining the behaviour of the static spin susceptibility
$\chi_s({\bf q},\omega=0)$ in the paramagnetic (PM) state: as the temperature
is lowered towards the critical temperature $T_{\mbox{\scriptsize{mag}}}$ of a
magnetically-ordered phase, the susceptibility will diverge at the
ordering vector ${\bf q}={\bf Q}$. Practically, we can implement this scheme
in a weak-coupling approach by first
calculating the phase diagram of the model using a two-site mean-field
ansatz, and then examining the location of the highest peaks in the
random-phase approximation (RPA) spin susceptibility at a temperature just
above $T_{\mbox{\scriptsize{mag}}}$ to check the 
validity of the low-temperature mean-field state. Clues to the origin
of the magnetic 
instability can be extracted from the dominant contributions to the
total static susceptibility from the orbitally resolved susceptibilities.

It is impossible to study each of the very many different orbital models
proposed for the pnictides, and so in this paper we will restrict ourselves to
five models: the two-orbital model of Raghu \emph{et al}.,~\cite{Raghu2008} the
three-orbital model of Daghofer \emph{et al}.,~\cite{Daghofer2010a} the
four-orbital model of Yu \emph{et al.},~\cite{Yu2009} and the five-orbital
models of Kuroki \emph{et al}.,~\cite{Kuroki2008} and Graser \emph{et
  al.}.~\cite{Graser2009} The main goal of this paper is to verify the
appearance and explore the 
robustness of the stripe AFM order in the weak-coupling limit in
these very different models, which 
have been selected as a representative sample of the different levels
of sophistication 
available in the literature. By this survey we not only hope to test the
suitability of these models and the weak-coupling theory to describe
the pnictide parent compounds, but 
also to gain insight into the origin of the AFM state and the features of the
electronic structure that
enhance or suppress it. We additionally hope to
identify an orbital model which 
gives the required AFM state and is also consistent with
key results 
of \emph{ab initio} calculations and ARPES data. Finally, the
role of Hartree 
shifts in renormalizing the band structure and the implications for
the magnetism is examined. 

Particular attention in our analysis will be paid to the
five-orbital models of Kuroki \emph{et al.} and Graser \emph{et al.} because
they are currently the most intensively-studied, 
the most sophisticated, and likely most accurate as they both result from
fits to \emph{ab initio} band structures of LaFeAsO. We note that other
five-orbital models of LaFeAsO were proposed in~\Ref{Calderon2009b}
and~\Ref{Ikeda2010}, but we will not examine them here. The latter is
very 
similar to Kuroki {\emph{et al.}}'s model, and therefore we neglect it here in
the interests of brevity. 
The former is quite different to Kuroki \emph{et al.}'s and Graser \emph{et
  al.}'s models as it was not the result of fitting to 
\emph{ab initio} calculations and does not include higher-order hopping
terms. This leads to very strong ellipticity of the electron Fermi pockets,
which substantially reduces Fermi-surface nesting and should
significantly alter the magnetic fluctuations. Since the focus of our
paper is on nesting and the resulting magnetic instabilities, we do
not consider this model.

The structure of the paper is as follows. In~\Sec{sec:theory} we outline the
theoretical basis of our analysis by discussing the general form of the
Hamiltonian, the mean-field decoupling scheme, and the construction of the
spin susceptibility within RPA. We then proceed to a systematic analysis of
the magnetic order in the five models in~\Sec{sec:results}. In
~\Sec{sec:discussion} we compare and contrast the ordering properties of the
models, and consider the implications for the pnictides. We
conclude with a summary in~\Sec{sec:summary}.

\section{Theory} \label{sec:theory} 

The general Hamiltonian for the orbital models is written $H = H_{0} + H_{I}$.
The non-interacting Hamiltonian $H_0$ is given by
\beq
H_{0} = \sum_{\bf k}\sum_{\sigma}\sum_{\nu,\mu}T_{\nu,\mu}({\bf
  k})d^{\dagger}_{{\bf k},\nu,\sigma}d^{}_{{\bf k},\mu,\sigma} \, ,
\eeq
where $d^{\dagger}_{{\bf k},\nu,\sigma}$ ($d^{}_{{\bf k},\nu,\sigma}$) is the
creation (annihilation) operator for a spin $\sigma$ electron of momentum
${\bf k}$ in the orbital $\nu$. $T_{\nu,\mu}({\bf k})$ are the kinetic
energy matrix elements for an effective tight-binding model defined in the
single-Fe unit cell of the two-dimensional
Fe-As planes.~\cite{Kuroki2008,Raghu2008,Graser2009,Yu2009,Daghofer2010a}

In all the proposed models, only on-site terms are included in the interaction
Hamiltonian 
\beqarray
H_{I} &=& U\sum_{{\bf i}}\sum_{\nu}n_{{\bf i},\nu,\uparrow}n_{{\bf
    i},\nu,\downarrow} + V\sum_{{\bf
    i}}\sum_{\nu\neq\mu}\sum_{\sigma,\sigma'}n_{{\bf i},\nu,\sigma}n_{{\bf
    i},\mu,\sigma'} \notag \\
&& - J\sum_{\bf i}\sum_{\nu\neq\mu}{\bf S}_{{\bf i},\nu}\cdot{\bf S}_{{\bf
      i},\mu} \nonumber \\
&&  + J'\sum_{\bf i}\sum_{\nu\neq\mu}d^{\dagger}_{{\bf
    i},\nu,\uparrow}d^{\dagger}_{{\bf i},\nu,\downarrow}d^{}_{{\bf
    i},\mu,\downarrow}d^{}_{{\bf i},\mu,\uparrow} \, . \label{eq:Hint}
\eeqarray
The number and spin operator for the orbital
$\nu$ at site ${\bf i}$ are defined as $n_{{\bf i},\nu,\sigma}=d^{\dagger}_{{\bf
    i},\nu,\sigma}d^{}_{{\bf i},\nu,\sigma}$ and ${\bf S}_{{\bf
    i},\nu}=\frac{1}{2}\sum_{\varsigma,\varsigma'}d^{\dagger}_{{\bf
    i},\nu,\varsigma}{\pmb \sigma}_{\varsigma,\varsigma'}d^{}_{{\bf
    i},\nu,\varsigma'}$, respectively, where ${\pmb
  \sigma}_{\varsigma,\varsigma'}$ is the vector of the Pauli spin
matrices. 
The matrix elements of the kinetic energy of all the models considered
here were
explicitly constructed to obey the orbital and lattice symmetries of
the Fe-As planes, which has been shown to be possible for a one-Fe unit cell
in 2-dimensional models.~\cite{Eschrig2009} Even though the interaction terms
are purely on-site, some 
restrictions on the parameters still arise from the requirement that it
should preserve the symmetries of the orbital space. The
orientation of the axes of the coordinate system used to 
define the orbitals must not change the symmetries of the Hamiltonian,
which implies $J = J'$ and $V = (2U - 5J)/4$.~\cite{Oles1983}

For the sake of brevity we only present results
for Hund's rule coupling $J/U=0.25$. Although this ratio is at the upper
limit of those 
employed in the literature, it is not unreasonable.
Screening can considerably reduce the effective
Coulomb interactions $U$ and $V$  from their atomic values, which
suggests rather large ratios of $J/U$. On the other hand, selecting $J \leq
U/3$ ensures that an electron added to an undoped site pays more energy to 
Coulomb repulsion than it can win from Hund's rule, i.e. that the
onsite interaction energy suppresses charge fluctuations rather than
enhancing them. Furthermore, a mean-field study of a number of models
has recently found that the  ordered moment and the reconstructed
Fermi surface in the $T=0$\,K AFM state is in good agreement with experiment
for our choice of $J/U$.~\cite{Luo2010}

\subsection{Mean-field analysis} \label{subsec:meanfield}

When constructing the mean-field phase diagrams of the orbital models, we
adopt the  frequently-employed 
ansatz~\cite{Yu2009,Bascones2010,Daghofer2010a,Daghofer2010b,Luo2010,Nomura2000}  
\beq
\langle n_{{\bf i},\nu,\sigma} \rangle = n_{\nu} + \frac{\sigma}{2}e^{i{\bf
    Q}\cdot{\bf r}_{\bf i}}m_{\nu} \label{eq:mfansatz} \, .
\eeq
Note that this neglects the possibility of inter-orbital 
averages $\langle{d^{\dagger}_{{\bf i},\nu,\sigma}d^{}_{{\bf
      i},\mu,\sigma}}\rangle$ ($\nu\neq\mu$), which 
have been included in some studies.~\cite{Kaneshita2009,Ran2009}
We restrict ourselves to ferromagnetic (FM) and AFM
phases with a two-site unit cell, corresponding to the ordering vectors ${\bf
  Q}=(0,0)$, ${\bf Q}=(\pi,0)$ and ${\bf 
  Q}=(\pi,\pi)$ in~\eq{eq:mfansatz}, respectively. The ${\bf Q}=(\pi,0)$
AFM state gives the magnetic order found in the pnictide parent
compounds.
Upon decoupling the interaction term
$H_I$, we obtain the mean-field Hamiltonian
\beqarray
H_{\mbox{MF}} & = &  H_0 + \sum_{\bf k}\sum_{\nu}\sum_{\sigma}\epsilon_{\nu}d^{\dagger}_{{\bf
    k},\nu,\sigma}d^{}_{{\bf k},\nu,\sigma} \notag \\
&& -\frac{1}{2}\sum_{\bf k}\sum_{\nu}\sum_{\sigma}\sigma\eta_{\nu}d^{\dagger}_{{\bf k}+{\bf
    Q},\nu,\sigma}d^{}_{{\bf k},\nu,\sigma} + NC \, ,\label{eq:mfham}
\eeqarray
where $N$ is the number of lattice sites and 
\beqarray
\epsilon_{\nu} & = & Un_{\nu} +
(2U-5J)\sum_{\mu\neq\nu}n_{\mu} \, , \label{eq:bareHshift} \\
\eta_{\nu} & = & Um_\nu + J\sum_{\mu\neq\nu}m_{\mu} \, , \label{eq:etanu} \\
C & = & -U\sum_{\nu}n_{\nu}^2 + \frac{U}{4}\sum_{\nu}m_{\nu}^2 -
(2U-5J)\sum_{\mu\neq\nu}n_{\nu}n_{\mu} \notag \\
&& + \frac{J}{4}\sum_{\mu\neq\nu}m_{\nu}m_{\mu} \, .
\eeqarray
Although the Hartree shifts $\epsilon_{\nu}$ can have rather large magnitude,
only the \emph{relative} shifts $\widetilde{\epsilon}_\nu = (5J-U)(n_\nu -
\min\{n_\mu\})$ are important when working at fixed doping. These relative
shifts are much smaller than the bare shifts $\epsilon_\nu$. Note that
not all mean-field studies include the Hartree
shifts.~\cite{Kaneshita2009,Ran2009} The influence of Hartree shifts
on the phase diagram will be discussed in~\Sec{subsec:HS}. 

For given temperature $T$ and
interaction constants $U$ and $J$, the stable mean-field state is determined
by the standard iterative diagonalization technique, where the condition of
constant charge density
\beq
\sum_{\nu}n_{\nu} = \begin{cases}
1 & \text{for 2 orbitals} \\
2 & \text{for 3 and 4 orbitals}\\
3 & \text{for 5 orbitals}
\end{cases}
\eeq
is enforced at each iteration step. Upon convergence, the free energy per
site is calculated. Comparing the free energy for each mean-field
ansatz allows us to determine the stable state of the system.

\subsection{The magnetic susceptibility}

We define the spin susceptibility by
\beqarray
\lefteqn{\chi^{jj'}({\bf q},i\omega_n)} \notag \\
& = & \frac{1}{N}\sum_{\nu,\mu}\int^{\beta}_{0}d\tau
e^{i\omega_n\tau}\left\langle T_{\tau} S^{j}_{\nu}({\bf
  q},\tau)S^{j'}_{\mu}(-{\bf q},0)\right\rangle \, ,
\eeqarray
where the Fourier-transformed spin operator $S^{j}_{\nu}({\bf q})$ is
related to the spin operator at site ${\bf i}$ by
\beq
S^{j}_{{\bf i},\nu} = \frac{1}{\sqrt{N}}\sum_{{\bf q}}S^{j}_{\nu}({\bf q}) e^{-i{\bf
    q}\cdot{\bf r}_{\bf i}} \, . \nonumber
\eeq
The total spin susceptibility is hence written as
\beq
\chi_{s}({\bf q},i\omega_n) = \sum_{j=x,y,z}\chi^{jj}({\bf q},i\omega_n) \, .
\eeq
To obtain the static spin susceptibilities we make the analytic continuation
$i\omega_n\rightarrow\omega+i0^+$ and then take the limit $\omega=0$. 
Note that it is not necessary to regularize the static susceptibility by
assuming a non-zero imaginary part of the frequency. Because
of the spin-rotation invariance of the PM state, we can express
$\chi_{s}({\bf q},\omega=0)$ in terms only of the static transverse
susceptibility 
\beq
\chi_{s}({\bf q},\omega=0) =
\frac{3}{2}\chi^{-+}({\bf q},\omega=0) \, .
\eeq
Henceforth we will only be concerned with $\chi^{-+}({\bf q},\omega=0)$
as this contains all relevant information on the magnetic ordering vector. 

We calculate the transverse spin susceptibility using the RPA. 
Introducing the generalized transverse susceptibilities 
\beqarray
\lefteqn{\chi^{-+}_{\nu,\nu',\mu,\mu'}({\bf q},i\omega_n)} \notag \\
& =& \frac{1}{N}\sum_{{\bf
    k},{\bf k}'}\int^{\beta}_{0}d\tau e^{i\omega_n\tau} \notag \\
&& \times\left\langle T_{\tau}
d^{\dagger}_{{\bf k}+{\bf q},\nu,\downarrow}(\tau)d^{}_{{\bf
    k},\nu',\uparrow}(\tau)d^{\dagger}_{{\bf k}'-{\bf
    q},\mu,\uparrow}(0)d^{}_{{\bf k}',\mu',\downarrow}(0)\right\rangle  \, ,
\notag \\
\eeqarray
we write the total transverse susceptibility 
\beq
\chi^{-+}({\bf q},i\omega_n) = \sum_{\nu,\mu}\chi^{-+}_{\nu,\nu,\mu,\mu}({\bf
  q},i\omega_n) \, . \label{eq:totchi}
\eeq
Summation of the ladder diagrams yields a Dyson equation for the generalized
susceptibilities 
\beqarray
\lefteqn{\chi^{-+}_{\nu,\nu',\mu,\mu'}({\bf q},i\omega_n)} \notag \\ &=&
\chi^{-+\,(0)}_{\nu,\nu',\mu,\mu'}({\bf q},i\omega_n) \notag \\
&& + \chi_{\nu,\nu',a,b}^{-+\,(0)}({\bf
  q},i\omega_n)V^{}_{a,b,c,d}\chi^{-+}_{c,d,\mu,\mu'}({\bf q},i\omega_n) \, , \label{eq:RPAeq}  
\eeqarray
where the non-zero elements of $V_{a,b,c,d}$ are
\begin{subequations}
\beqarray
V_{a,a,a,a} & = & U \, ,\\ 
V_{a,a,b,b} & = & J \, ,\\
V_{a,b,b,a} & = & U -2J \, , \\
V_{a,b,a,b} & = & J \, ,
\eeqarray
\end{subequations}
and we assume $a\neq b$.
The mean-field susceptibilities $\chi^{-+\,(0)}_{\nu,\nu',\mu,\mu'}({\bf
  q},i\omega_n)$ can be explicitly written as
\beqarray
\lefteqn{\chi^{-+\,(0)}_{\nu,\nu',\mu,\mu'}({\bf q},i\omega_n)} \notag \\
& = & -\frac{1}{N}\sum_{\bf
  k}\sum_{s,s'}u^{}_{s,\nu'}({\bf k})u^{\ast}_{s,\mu}({\bf
  k})u^{}_{s',\mu'}({\bf k}+{\bf q})u^{\ast}_{s',\nu}({\bf k}+{\bf q}) \notag \\
&& \times\frac{n_{F}(E_{s,{\bf k}}) - n_{F}(E_{s',{\bf k}+{\bf q}})}{E_{s,{\bf
      k}} - E_{s',{\bf k}+{\bf q}} - i\omega_n} \, ,
\eeqarray
where $n_{F}(E)$ is the Fermi function, $E_{s,{\bf k}}$ are the eigenvalues
of $H_{\mbox{\scriptsize{MF}}}$~\eq{eq:mfham}, and the coefficients
$u_{s,\nu}({\bf k})$ transform the 
diagonalizing annihilation operators $\gamma^{}_{s,{\bf k}}$ corresponding to
the eigenvalues $E_{s,{\bf k}}$ into the orbital
basis, i.e. $d^{}_{\nu,{\bf k}} = \sum_{s}u_{s,\nu}({\bf k})\gamma^{}_{s,{\bf
    k}}$.  

Apart from the total transverse spin susceptibility, it is also interesting to
consider the dominant contributions to the sum~\eq{eq:totchi}. It is
convenient to define
$\chi^{-+}_{\nu,\mu}({\bf q},i\omega_n) \equiv \chi^{-+}_{\nu,\nu,\mu,\mu}({\bf
  q},i\omega_n)$. These orbitally resolved susceptibilities can provide
important insight into the 
regions of the Fermi surface most strongly involved in the magnetic
order. We adopt the following enumeration of the orbitals: $\nu=1$
($xz$), $\nu=2$ ($yz$), $\nu=3$ ($xy$), $\nu=4$ ($x^2-y^2$), and $\nu=5$
($3z^2-r^2$).

\section{Results} \label{sec:results}

In this section we present a systematic analysis of the magnetic
behaviour of the two-orbital model of Raghu \emph{ et 
  al}.~\cite{Raghu2008} (\Sec{subsec:Raghu}), the three-orbital model of
Daghofer \emph{et 
  al.}~\cite{Daghofer2010a} (\Sec{subsec:Daghofer}), the four-orbital model of
Yu \emph{et 
  al.}~\cite{Yu2009} (\Sec{subsec:Yu}), and the five-orbital models of Kuroki
\emph{et 
  al.}~\cite{Kuroki2008} (\Sec{subsec:Kuroki}) and Graser \emph{et
  al.}~\cite{Graser2009} (\Sec{subsec:Graser}). In all cases, we first
determine the mean-field phase diagram using the 
method outlined in~\Sec{subsec:meanfield}, and then calculate the
susceptibilities at several points in the PM state lying close to the
boundary of the ordered phases. We emphasize that we do not seek to
  construct the true mean-field 
phase diagram of the orbital models, but rather to determine the
appropriateness of the mean-field ansatz~\eq{eq:mfansatz}.

All calculations (i.e. the phase diagram and the mean-field susceptibilities)
were performed using a $400\times400$ ${\bf k}$-point mesh. In order to better
distinguish the total and the orbitally resolved
susceptibilities from one another, we use a different color scheme in the
density plots of these quantities. We also note that the total static
susceptibility is symmetric about the line $q_x=q_y$; if peaks in
$\chi^{-+}({\bf q},\omega=0)$ are found off this line, for simplicity we will
only mention the peak with $q_x>q_y$ in the discussion.

\subsection{The two-orbital model of Raghu \emph{et al.}} \label{subsec:Raghu}

\begin{figure}
  \begin{center}
  \includegraphics[clip,width=\columnwidth]{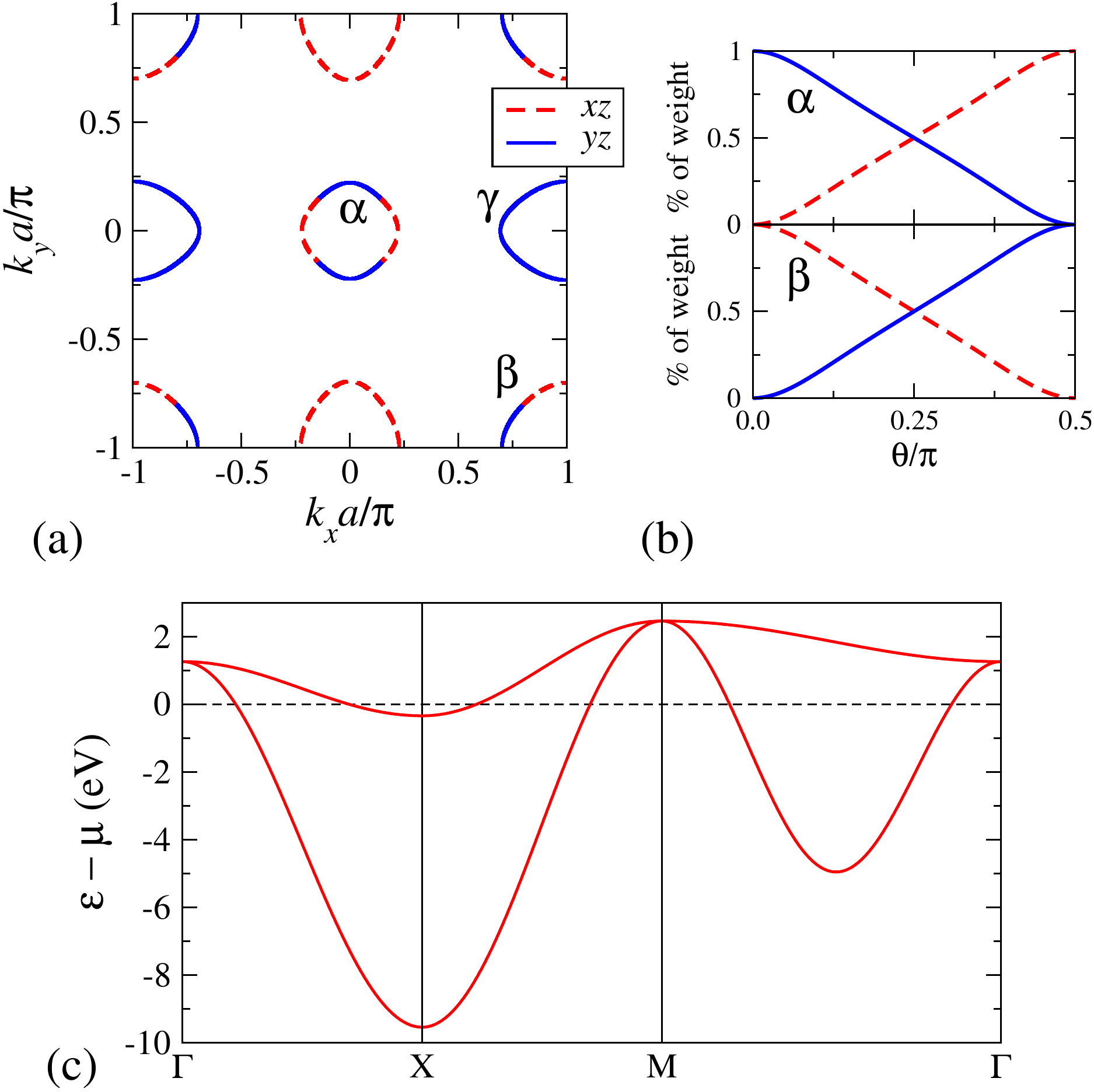}
  \end{center}
  \caption{\label{2b_FS_BS} (a) The Fermi surface of the
    two-orbital model of Raghu \emph{et al.}~\cite{Raghu2008} showing the
    dominant orbital contribution. (b) The 
    orbital weight around the $\alpha$ and $\beta$ Fermi surfaces as a
    function of the winding angle $\theta$ measured with respect to the
    $y$-axis, taken in the anti-clockwise direction. The orbital 
    weight around the 
    $\gamma$ Fermi surface is not shown as it has almost complete $yz$
    character. (c) The 
    band structure along high-symmetry directions.}  
\end{figure}

The two-orbital model of Raghu \emph{et al.}~\cite{Raghu2008} was one of the
first attempts to model the pnictides from an orbital point of view, and
also the first of a number of similar two-orbital
models.~\cite{Daghofer2008,Ran2009,Moreo2009,Calderon2009a}
As it only keeps the $xz$ and $yz$ orbitals, it may be regarded as a
minimal orbital model of the system. Although it is likely too simple to give
quantitative 
agreement with experiment, it is nevertheless of interest to us as it
qualitatively captures the variation of the orbital character about the Fermi
surface and hence provides important clues to the ordering mechanism.

The Fermi surface and the band structure of the two-orbital model is shown
in~\fig{2b_FS_BS}.~\cite{2bandnote} As we shall see, the Fermi
surface is somewhat unlike those of the more sophisticated models, specifically
one of the $xz$/$yz$-derived hole pockets is located at
$(\pi,\pi)$ instead of at $(0,0)$. This implies rather different magnetic
properties compared to these other models, as the nesting of the
hole pocket at $(\pi,\pi)$ with the electron pocket at $(0,\pi)$ allows
both electron pockets to participate on an almost equal footing in the
$(\pi,0)$ AFM state.

\begin{figure}
  \begin{center}
  \includegraphics[clip,width=\columnwidth]{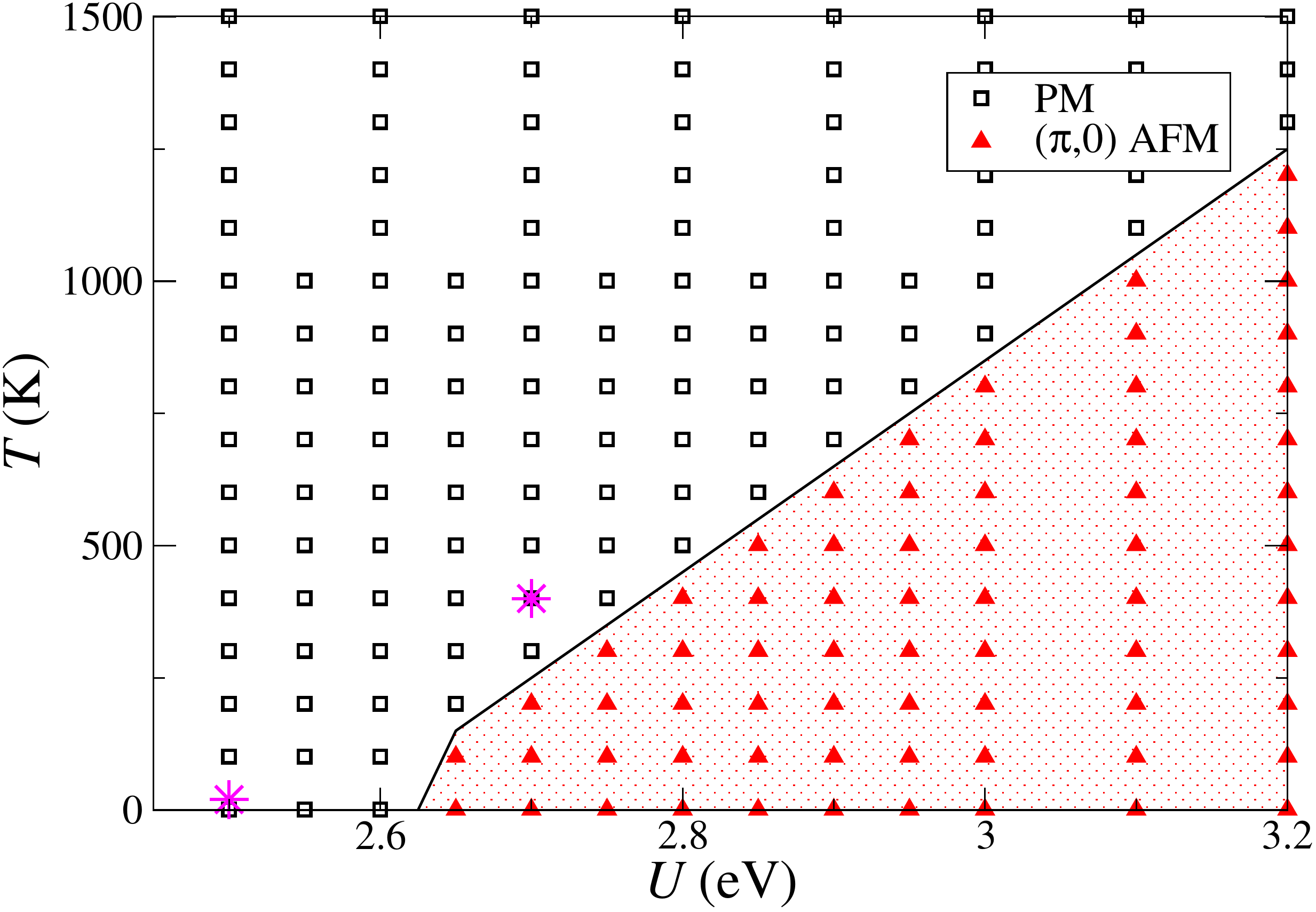}
  \end{center}
  \caption{\label{mf_2b} The $T$-$U$ mean-field phase diagram of
    the two-orbital model of Raghu \emph{et al.}.~\cite{Raghu2008} The location
    of the symbols indicate a co-ordinate at 
    which the free energy was evaluated, and the corresponding minimizing
    state. The shaded region approximately indicates the extent
    of the $(\pi,0)$ AFM phase. We show the static susceptibility at the points
    indicated by the star symbols in~\fig{chi_2b}.}  
\end{figure}

The mean-field phase diagram of the two-orbital model is presented
in~\fig{mf_2b}. The $(\pi,0)$ AFM state is 
stable above a critical interaction strength
$U_c\approx2.65$\,eV, with the critical temperature increasing linearly with $U$
above this value. Our results are consistent with those of~\Ref{Kubo2009} for
$J=0.1U$.  
In a similar two-orbital model,~\cite{Lorenzana2008} the mean-field
ground state was found to be a superposition of ${\bf Q}_x=(\pi,0)$ and ${\bf
  Q}_{y}=(0,\pi)$ stripe magnetic orders with mutually-perpendicular spin polarizations. Although our aim is to check the validity of the
mean-field phase diagram by examining the spin susceptibility just above the
mean-field ordering temperature, we are not able to determine whether such a
two-${\bf Q}$ state or the expected $(\pi,0)$ AFM phase is realized: in both
cases we will find a peak in the susceptibility at $(\pi,0)$. 

\begin{figure*}
  \begin{center}
  \includegraphics[clip,width=2\columnwidth]{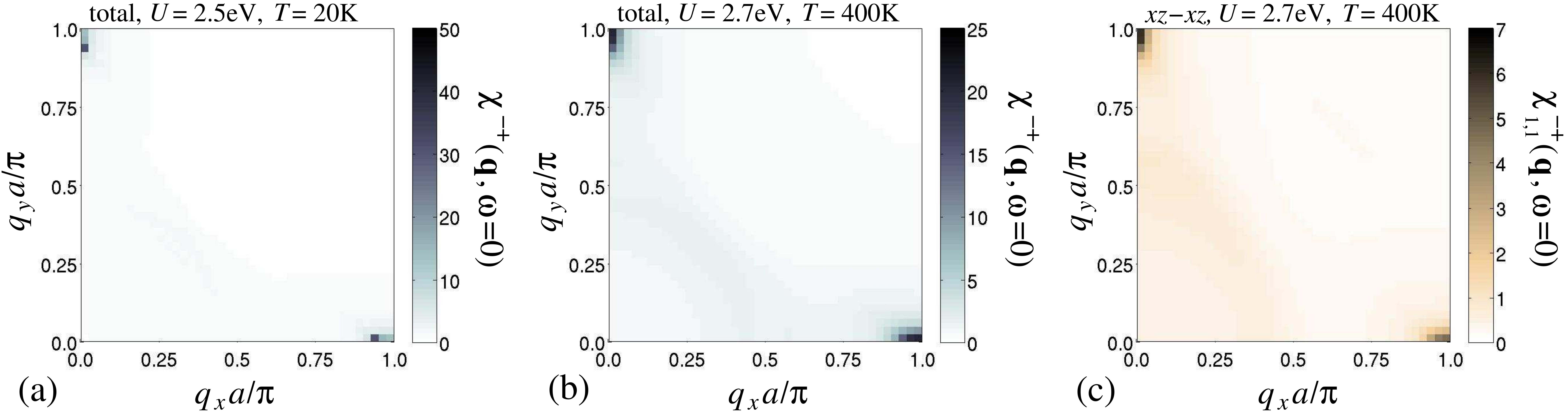}
  \end{center}
  \caption{\label{chi_2b} The static transverse spin
    susceptibility of the two-orbital model of Raghu \emph{et al.}~\cite{Raghu2008} at
    (a) $U=2.5$\,eV and $T=20$\,K and (b) $U=2.7$\,eV and $T=400$\,K. (c) The
    dominant contribution to the static
    susceptibility from the $xz$-$xz$ term at $U=2.7$\,eV
    and $T=400$\,K. Note the different scales used in each panel.}   
\end{figure*}

In~\fig{chi_2b} we show the static susceptibility at the points indicated
in~\fig{mf_2b}. At both points 
the susceptibility is sharply peaked at or very close to $(\pi,0)$, 
reflecting the good nesting of the electron and hole pockets, and confirming the
appropriateness of our mean-field ansatz. The absence of
any other significant structure in the susceptibility implies that only scattering
between the electron and hole pockets is important to the magnetic
ordering. Examining the susceptibility at $\approx150$\,K above the AFM
transition temperature at a number of interaction strengths between $U=2.7$\,eV
and $3$\,eV, we 
find only minor quantitative differences in $\chi^{-+}({\bf q},\omega=0)$
within this range.
At $U<U_c$ the magnetic response indicates weak incommensuration, with
the susceptibility peaking at ${\bf q}\approx(0.95\pi,0)$
[see~\fig{chi_2b}(a)]. An
incommensurate AFM phase is in fact realized at $U=2.6$\,eV and $T=100$\,K, as
revealed by the observation of negative values of $\chi^{-+}({\bf
  q},\omega=0)$ here. 
The unrestricted real-space Hartree-Fock of~\Ref{Lorenzana2008} may have
missed the weakly-incommensurate phase because the considered cluster
size was too small. 

The dominant contribution to the static transverse susceptibility is from the
$xz$-$xz$ and $yz$-$yz$ terms. $\chi^{-+}_{1,1}({\bf q},\omega=0)$ is shown
in~\fig{chi_2b}(c); $\chi^{-+}_{2,2}({\bf q},\omega=0)$ can be obtained from
this by interchanging $q_x$ and $q_y$. We see that although there are definite
peaks at both $(\pi,0)$ and $(0,\pi)$, the latter peak is somewhat higher than
the former. This implies that the $yz$ [$xz$] orbital is slightly more
important to $(\pi,0)$ [$(0,\pi)$] AFM order, consistent with the mean-field
calculation in~\Ref{Kubo2009}. 

\subsection{The three-orbital model of Daghofer \emph{et al.}} \label{subsec:Daghofer}

\begin{figure}
  \begin{center}
  \includegraphics[clip,width=\columnwidth]{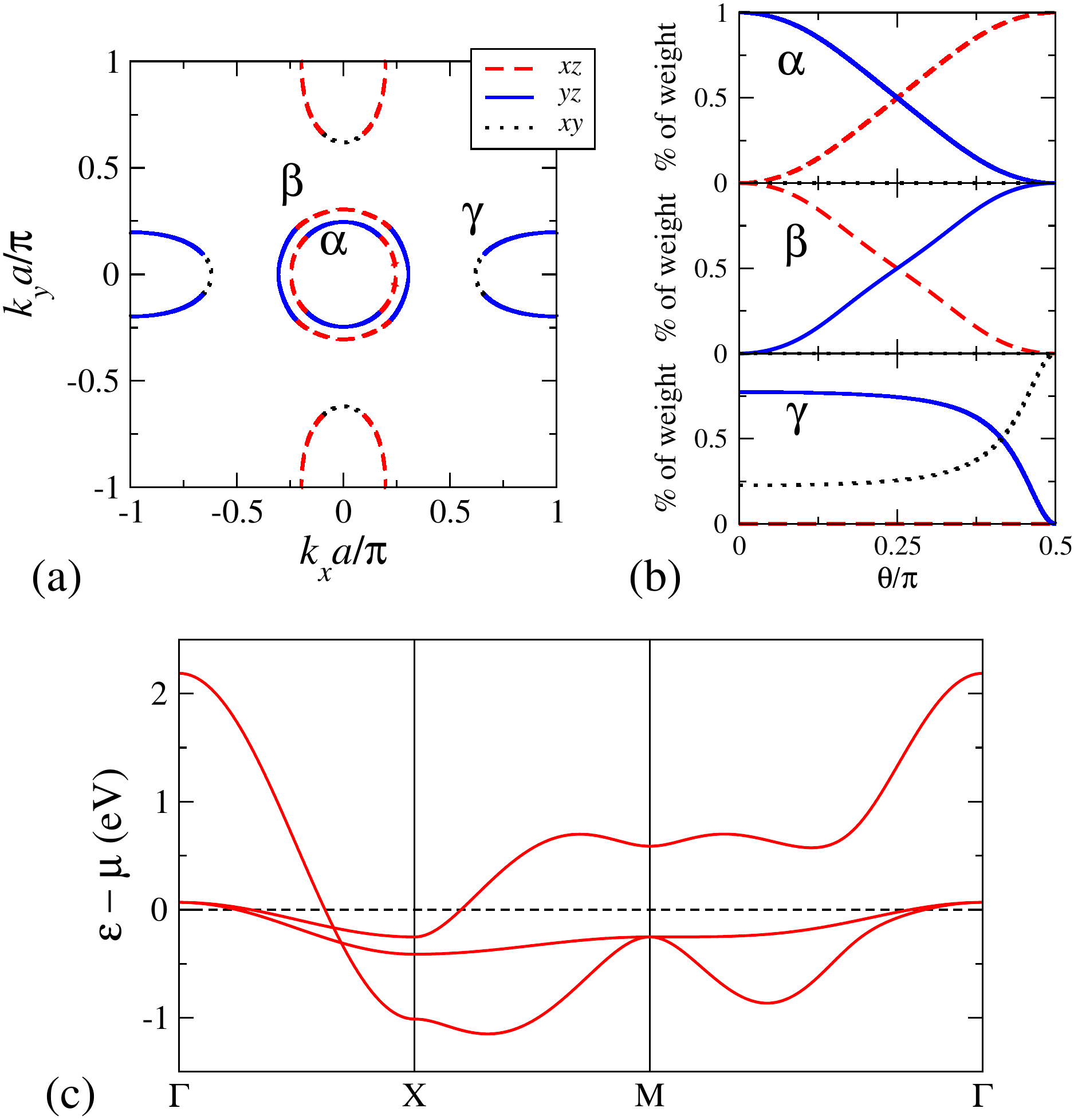}
  \end{center}
  \caption{\label{3b_FS_BS} (a) The Fermi surface of the
    three-orbital model of Daghofer \emph{et al.}~\cite{Daghofer2010a} showing the dominant orbital
    contribution. (b) The 
    orbital weight around the labeled Fermi surfaces. (c) The 
    band structure along high-symmetry directions.}  
\end{figure}

\begin{figure}
  \begin{center}
  \includegraphics[clip,width=\columnwidth]{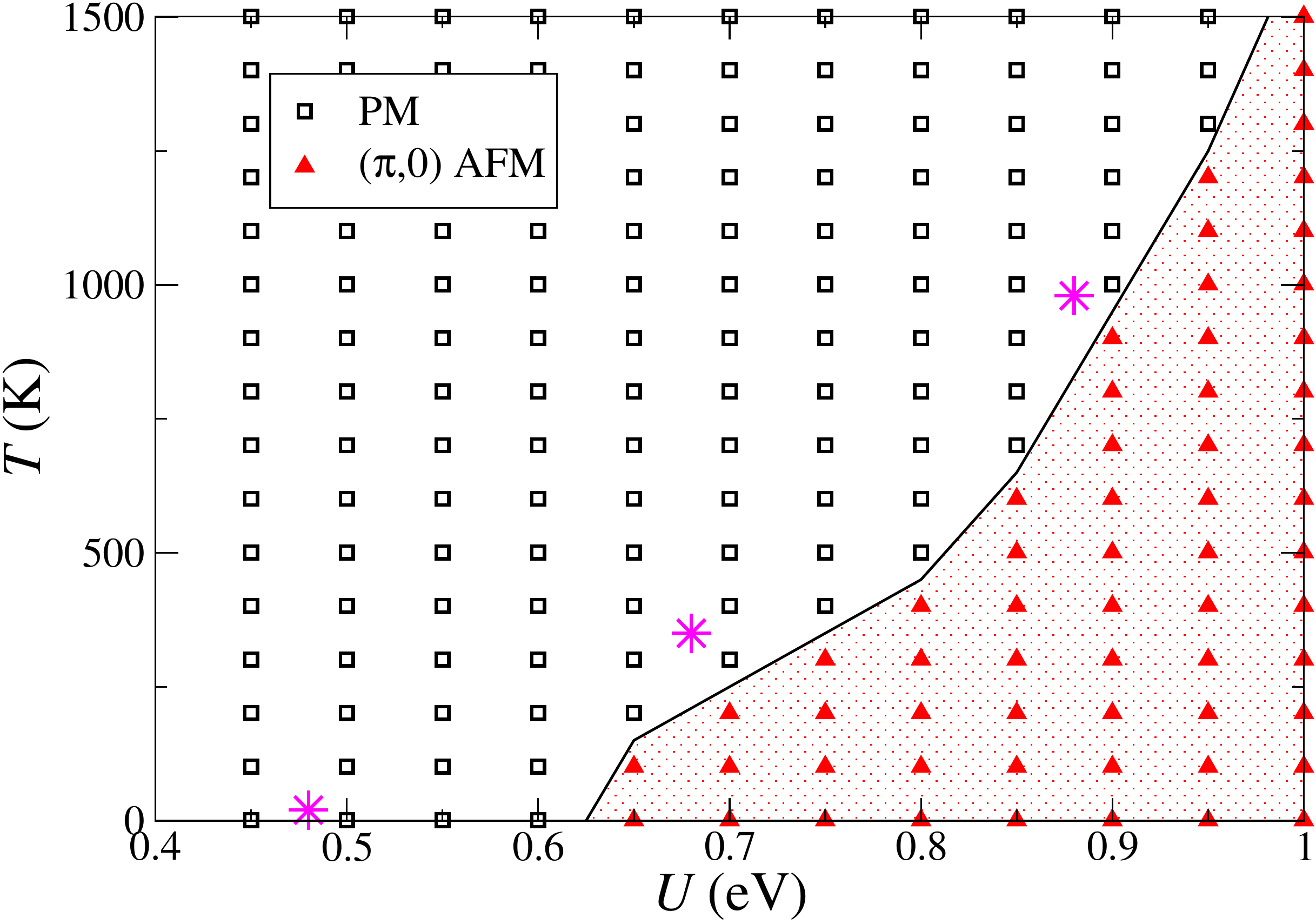}
  \end{center}
  \caption{\label{mf_3b} The $T$-$U$ mean-field phase diagram of
    the three-orbital model of Daghofer \emph{et al.}.~\cite{Daghofer2010a}.
    The shaded region approximately indicates the 
    extent of the $(\pi,0)$ AFM phase. We show the static susceptibility at
    the points 
    indicated by the star symbols in~\fig{chi_3b}.}  
\end{figure}

The three-orbital model of Daghofer \emph{et al.}~\cite{Daghofer2010a} is the
newest model studied in this work. In addition to the $xz$ and $yz$ orbitals,
this model also includes the $xy$ orbital. Assuming that the magnetic order is
driven by nesting of electron and hole Fermi surfaces, it is reasonable to
neglect the $x^2-y^2$ and $3z^2-r^2$ orbitals as \emph{ab initio} calculations
predict that they have 
very little weight at the Fermi energy.~\cite{Boeri2008,Eschrig2009}

The electronic structure of the three-orbital model is summarized
in~\fig{3b_FS_BS}. The model reproduces key features of the Fermi surface
predicted by \emph{ab initio} calculations: there are two $xz/yz$-derived
nearly circular hole pockets at the 
centre of the Brillouin zone, and a $yz/xy$-derived [$xz/yz$-derived]
elliptical electron pocket centred at ${\bf k}=(\pi,0)$ [${\bf
    k}=(0,\pi)$]. We note that the hole pockets have much greater effective
mass than the electron pockets, in striking contrast to the two-orbital model
discussed above.

The ground state phase diagram of the model as a function of $J$ and $U$ was
mapped in~\Ref{Daghofer2010a} using a mean-field ansatz
that also allowed for stripe or staggered orbital order in addition to FM and
AFM states. At
moderate- to strong-coupling, orbital ordered states were obtained; in the
weak-coupling regime and $J=0.25U$, however, the $(\pi,0)$ AFM state without
orbital order was found to be stable. The authors of~\Ref{Luo2010} mapped the
same phase diagram using a more restricted mean-field ansatz, but their RPA
calculations suggest that the $(\pi,0)$ AFM state is unstable towards
incommensurate order. In~\fig{mf_3b}, we show the phase diagram in the $T$-$U$
plane.

\begin{figure*}
  \begin{center}
  \includegraphics[clip,width=2\columnwidth]{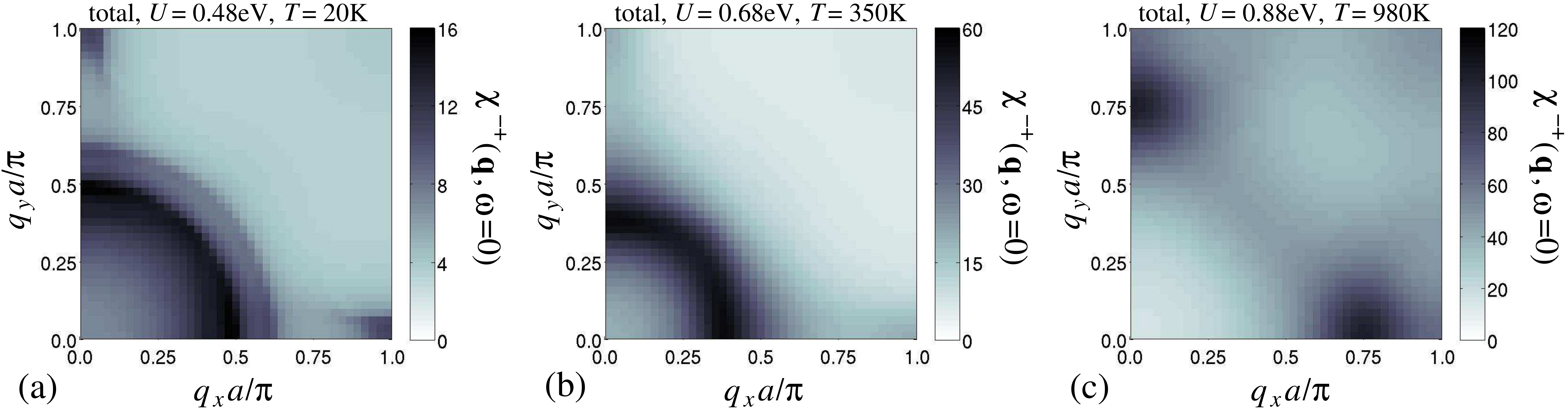}
  \end{center}
  \caption{\label{chi_3b} The static transverse spin
    susceptibility of the three-orbital model of Daghofer \emph{et
      al.}~\cite{Daghofer2010a} at 
    (a) $U=0.48$\,eV and $T=20$\,K, (b) $U=0.68$\,eV and $T=350$\,K, and (c)
    $U=0.88$\,eV and $T=980$\,K.}  
\end{figure*}

Our results for the spin susceptibility are in agreement
with~\Ref{Luo2010}, and we therefore conclude that the phase
diagram~\fig{mf_3b} does not  
reflect the actual mean-field behaviour of the model. At relatively weak
coupling the dominant magnetic response occurs in a ring of almost uniform
height centred at ${\bf q}=(0,0)$ of radius $\approx0.5\pi$ and $0.4\pi$,
see~\fig{chi_3b}(a) and (b) respectively.  
In the former case, we can make out 
another, fainter ring of radius $\approx0.65\pi$; the radii of these
rings correspond exactly to twice the average radii of the hole
pockets, indicating that the magnetic response is dominated by
scattering from one side of the hole Fermi surfaces to the other. The
smaller radius of the ring in the $U=0.68$\,eV data reflects the reduction of
the size of the heavy hole pockets by the increase of the chemical potential
with temperature.  
In both of these figures, there is only a very small peak close to
$(\pi,0)$; 
this peak is barely visible in the $U=0.68$\,eV, $T=350$\,K data.  
The form of the susceptibility changes significantly at higher $U$: as shown
in~\fig{chi_3b}(c), two rather broad peaks develop at ${\bf
  q}\approx(0.75\pi,0)$. The origin of this behaviour is
unclear, as this wavevector does not seem to correlate with any Fermi surface
or band structure feature.

It is interesting to consider the consequence of shrinking the hole
Fermi surfaces in this model, for example by electron-doping. We expect that
this should also
shrink the ring feature in the susceptibility, leading to significant weight
close to ${\bf q}=0$, favouring FM order and triplet
superconductivity. Such a situation might be relevant to the stochiometric
superconductor LiFeAs, where ARPES
reveals small heavy hole pockets which are poorly nested with electron
Fermi surfaces.~\cite{LiFeAsARPES} A model for this compound which captures
these salient band structure features was recently shown to
display FM order and triplet superconductivity in the weak-coupling
limit.~\cite{Brydon2011} Intriguingly,  
unpublished NMR results on single crystals of LiFeAs report that there is no
suppression of the Knight shift below $T_{c}$ for magnetic fields
perpendicular to the crystal $c$ axis, strongly suggesting triplet
pairing.~\cite{LiFeAsNMR}

We note that there are several alternative three-orbital models
which display an $xy$-dominated hole pocket at the M
point.~\cite{Lee2008,SLYu2009} A similar analysis to that performed here has
been carried out for such a three-orbital model,~\cite{Long2010} and the
susceptibility was found to be peaked at an incommensurate wavevector close
to $(\pi,0)$. More recently, Zhou and Wang studied a three-orbital model
of LaFeAsO using Gutzwiller mean-field theory restricted to two-site magnetic
unit cells and found significant deviations from the usual Hartree-Fock
approach.~\cite{Zhou2010}

\subsection{The four-orbital model of Yu \emph{et al.}} \label{subsec:Yu}

\begin{figure}
  \begin{center}
  \includegraphics[clip,width=\columnwidth]{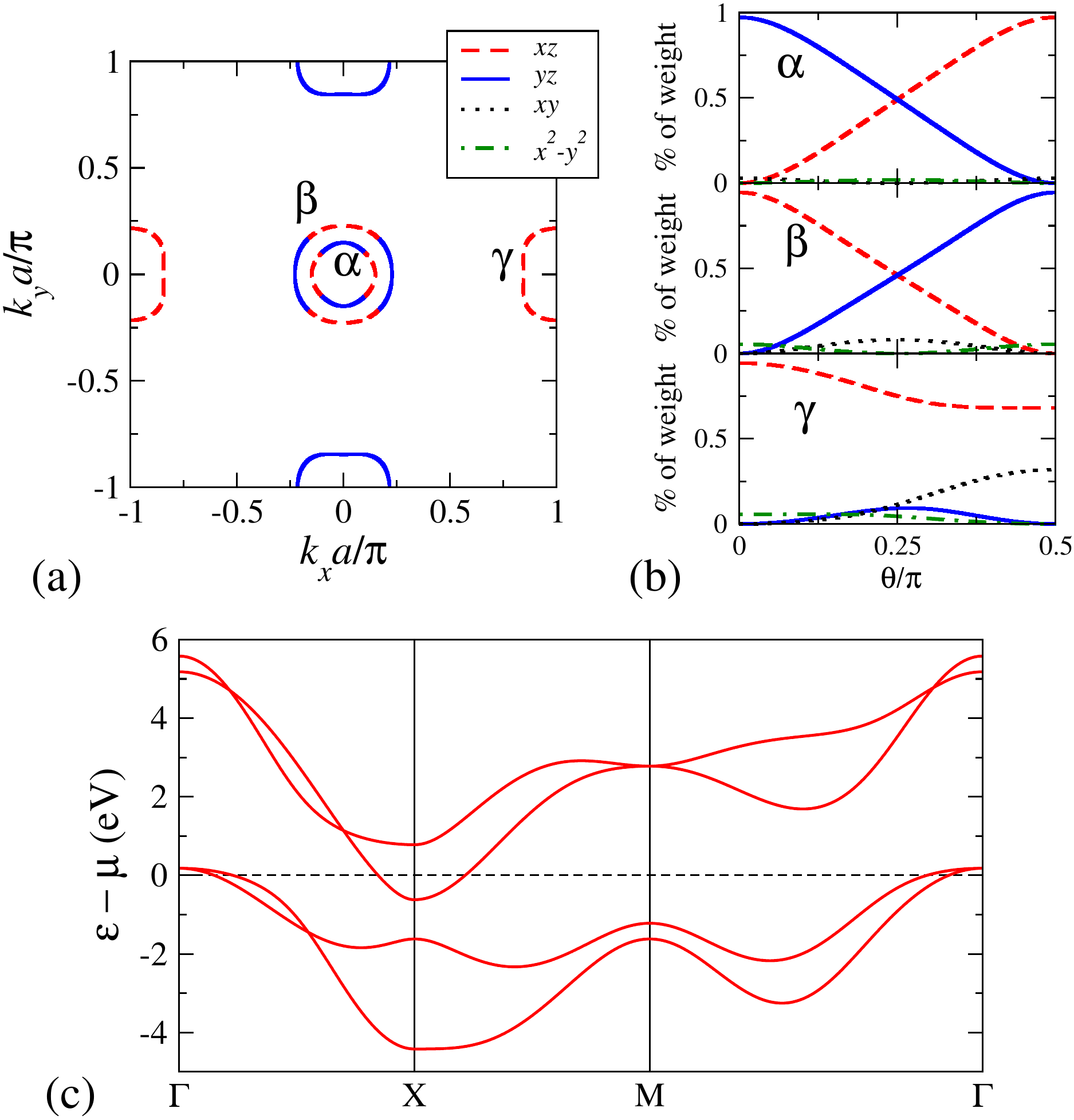}
  \end{center}
  \caption{\label{4b_FS_BS} (a) The Fermi surface of the
    four-orbital model of Yu \emph{et al.}~\cite{Yu2009} showing the dominant
    orbital contribution. (b) The 
    orbital weight around the labeled Fermi surfaces in (a). (c) The
    band structure along high-symmetry directions.}  
\end{figure}

\begin{figure}
  \begin{center}
  \includegraphics[clip,width=\columnwidth]{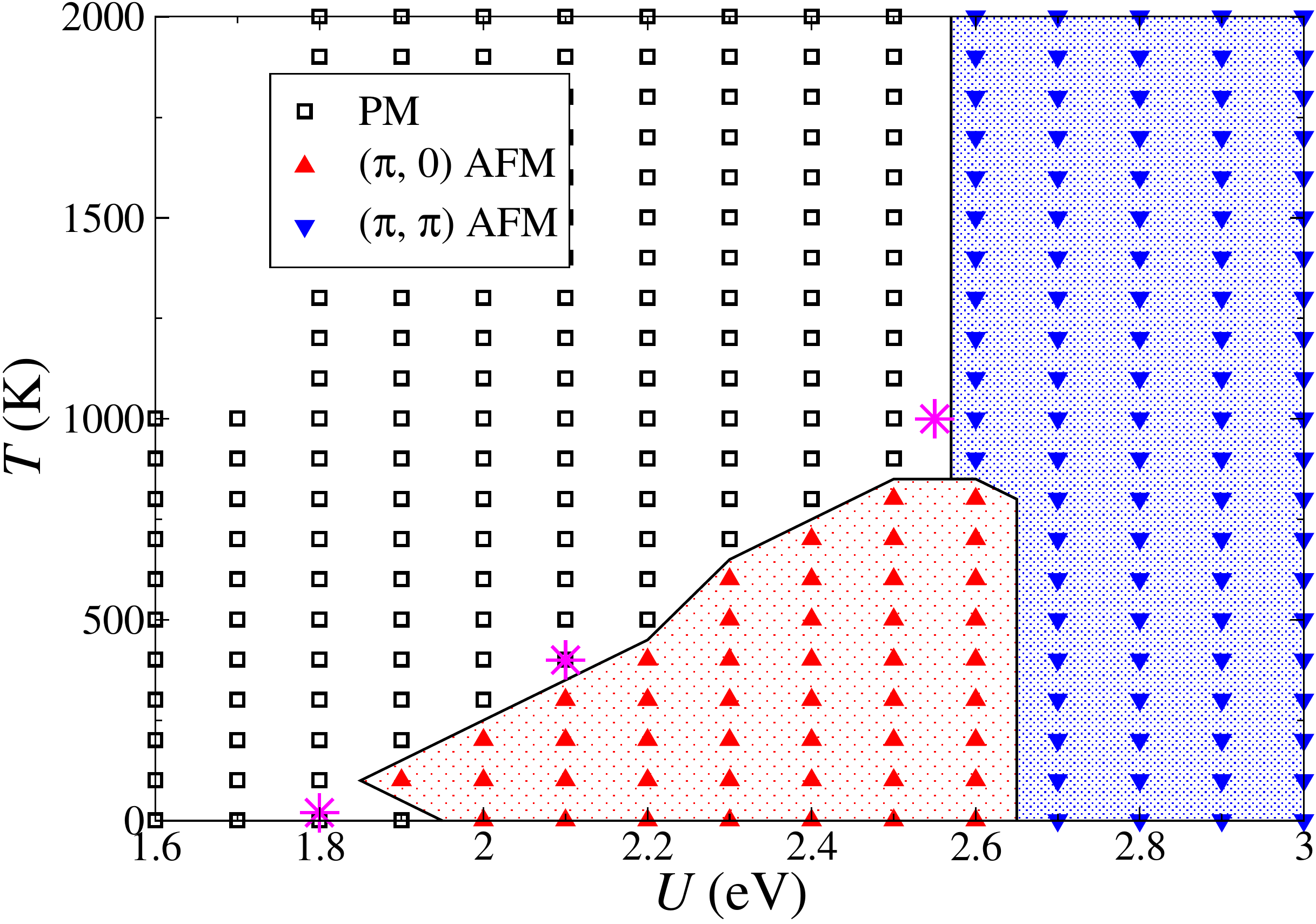}
  \end{center}
  \caption{\label{mf_4b} The $T$-$U$ mean-field phase diagram of
    the four-orbital model of Yu \emph{et al.}.~\cite{Yu2009} The
    lightly-shaded region approximately indicates 
    the 
    extent of the $(\pi,0)$ AFM phase, while the darkly-shaded region gives
    the extent of the $(\pi,\pi)$ AFM phase. We show the static susceptibility
    at the points indicated by the star symbols in~\fig{chi_4b}.}  
\end{figure}

The four-orbital model of Yu \emph{et al.}~\cite{Yu2009} includes all but the
$3z^2-r^2$ orbital. Although neither the $x^2-y^2$ or the $3z^2-r^2$ orbitals
significantly contribute to the Fermi surface, the former is filled at
  lower doping levels. This suggests that the
$3z^2-r^2$ orbital will be less relevant to the magnetism,~\cite{Arita2009},
justifying its exclusion from this model. 

As shown in~\fig{4b_FS_BS}, the
structure of the Fermi surface is broadly similar to that of the three-orbital
model, with elliptical electron Fermi pockets at the X points, and two nearly
circular  
hole pockets with much greater effective mass at the $\Gamma$ point.
Note, however, that the electron pockets are rotated and have reversed
dominant orbital content (i.e. $xz \leftrightarrow yz$), while each Fermi
surface is smaller. These are likely relatively superficial differences:
the orientation of the electron pockets is correct in the (physical) two-Fe
Brillouin zone, while the size of the Fermi
pockets can be tuned through the tight-binding parameters. More seriously, 
however, the model contradicts some key predictions of \emph{ab initio}
calculations: the $xy$ weight at the Fermi surface is heavily
suppressed,~\cite{Boeri2008,Eschrig2009} and 
only a single band contributes to the electron Fermi
pockets.~\cite{nesting,Boeri2008}

Our mean-field phase diagram agrees with the $T=0$\,K analysis of Yu \emph{et
  al.}: we find a low-temperature $(\pi,0)$ AFM state for
$1.9\mbox{eV}\lesssim U \lesssim 2.65\mbox{eV}$, while above a 
critical $U_{c}\approx2.65$\,eV the ground state is the $(\pi,\pi)$ AFM
  which has a very 
high ordering temperature ($>3000$\,K). It is possible that the strong-coupling
$(\pi,0)$ AFM state found in~\Ref{Yu2009} has lower free energy
for $U>4$\,eV, but at these coupling strengths the AFM states are insulating and
hence irrelevant to the magnetic ordering in the pnictides. 

\begin{figure*}
  \begin{center}
  \includegraphics[clip,width=2\columnwidth]{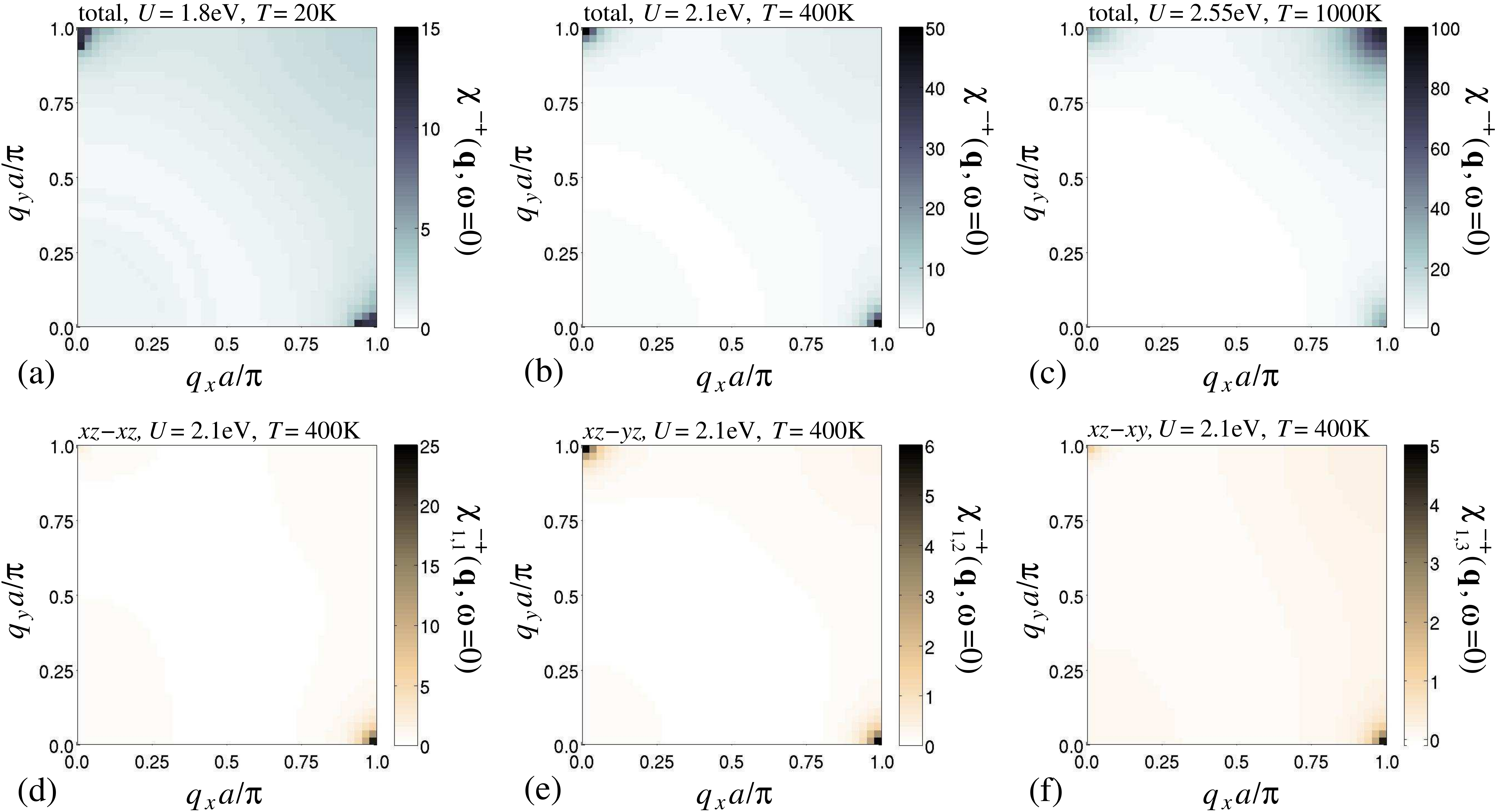}
  \end{center}
  \caption{\label{chi_4b} The static transverse spin
    susceptibility of the four-orbital model of Yu \emph{ et
      al.}~\cite{Yu2009} at
    (a) $U=1.8$\,eV and $T=20$\,K, (b) $U=2.1$\,eV and $T=400$\,K, and (c) $U=2.55$\,eV
    and $T=1000$\,K. The remaining three figures show the dominant contributions
    to the peak at ${\bf q}=(\pi,0)$ at $U=2.1$\,eV and $T=400$\,K: (d) the
    $xz$-$xz$ term, (e) the $xz$-$xy$ term, and (f) the $xz$-$xy$ term.}  
\end{figure*}

In~\fig{chi_4b} we show the static transverse susceptibility at the points
indicated in the phase diagram~\fig{mf_4b}. The low-$U$ results [\fig{chi_4b}(a) and (b)] are strongly reminiscent of the
two-orbital model. For $U=1.8$\,eV and $T=20$\,K, just below the critical coupling
strength of the $(\pi,0)$ AFM state, $\chi^{-+}({\bf q},\omega=0)$ is peaked
at ${\bf q}\approx(0.95\pi,0)$. Closer to the $(\pi,0)$ AFM phase
there is evidence that an ordered incommensurate state is realized. At
$U=2.1$\,eV and $T=400$\,K, we find a sharp peak exactly at ${\bf
  q}=(\pi,0)$. These peaks are clearly derived from  
the very good nesting of the electron and hole Fermi surfaces. This
conclusion is supported by examination of the dominant contributions to the
peak at $(\pi,0)$ in~\fig{chi_4b}(d-f): the $xz$-$xz$ term has the
largest value, but the sum of the $xz$-$yz$ and $xz$-$xy$ terms is
comparable. This 
is to be expected, as the $xz$ orbital dominates the electron pocket at ${\bf
  k}=(\pi,0)$ [see~\fig{4b_FS_BS}]; furthermore, the elliptical form of the
electron pocket means that it is nested with the $xz$-dominated regions of the
hole pockets, hence accounting for the dominance of
  $\chi^{-+}_{1,1}$. There 
is nevertheless significant $yz$ and $xz$ orbital weight 
around the hole and electron pockets, respectively, accounting for the
importance of the inter-orbital susceptibilities $\chi^{-+}_{1,2}$ and
$\chi^{-+}_{1,3}$. 

Consistent with the mean-field phase diagram, a broad peak develops in
$\chi^{-+}({\bf q},\omega=0)$ at ${\bf 
  q}=(\pi,\pi)$ with increasing $U$; close to the $(\pi,\pi)$ AFM state it
dominates the magnetic response, see~\fig{chi_4b}(c).
Although there is low-energy scattering between the two electron
pockets with this wavevector, the large magnitudes of the mean-field
staggered magnetic potentials $\eta_{\nu}$ [\eq{eq:etanu}] suggests that 
the $(\pi,\pi)$ AFM state cannot be understood only in terms of Fermi surface
physics. 
For example, at $U=2.7$\,eV and $3$\,eV we have 0.53\,eV$\geq|\eta_{\nu}|\geq0.43$\,eV
and $1.23$\,eV$\geq|\eta_\nu|\geq0.93$\,eV, 
respectively. The former is comparable to the minimum energy (at the X points)
of the band lying entirely above the Fermi surface; the latter is close to the
difference between the hole bands and the Fermi energy at the M point. 
Since the $\eta_\nu$ are the characteristic energies for the 
reconstruction of the electronic structure in the AFM state,  it is
therefore likely that these regions participate in the $(\pi,\pi)$ AFM
state, accounting for its much lower free energy compared to the $(\pi,0)$ AFM 
phase.~\cite{Yu2009}

\subsection{The five-orbital model of Kuroki \emph{et al.}} \label{subsec:Kuroki}

The five-orbital model of Kuroki \emph{et al.}~\cite{Kuroki2008} includes all
five Fe $d$-orbitals. It was constructed by fitting to \emph{ab initio}
calculations for LaFeAsO,
and it was one of the first orbital-based models proposed for the iron
pnictides. The behaviour of this model, and several of its close
relatives, has been extensively 
studied;~\cite{Ikeda2008,Kaneshita2009,Kariyado2009,Arita2009,Kuroki2009,Kaneshita2010,Ikeda2010}
the magnetic properties are particularly well understood, with RPA calculations
confirming that $(\pi,0)$ AFM order is realized in the undoped
model.~\cite{Kariyado2009,Kaneshita2010} The extent 
of this state in the phase diagram at zero doping, and the possible
competition with other 
magnetic phases, nevertheless remains largely unexplored.

\begin{figure}
  \begin{center}
  \includegraphics[clip,width=\columnwidth]{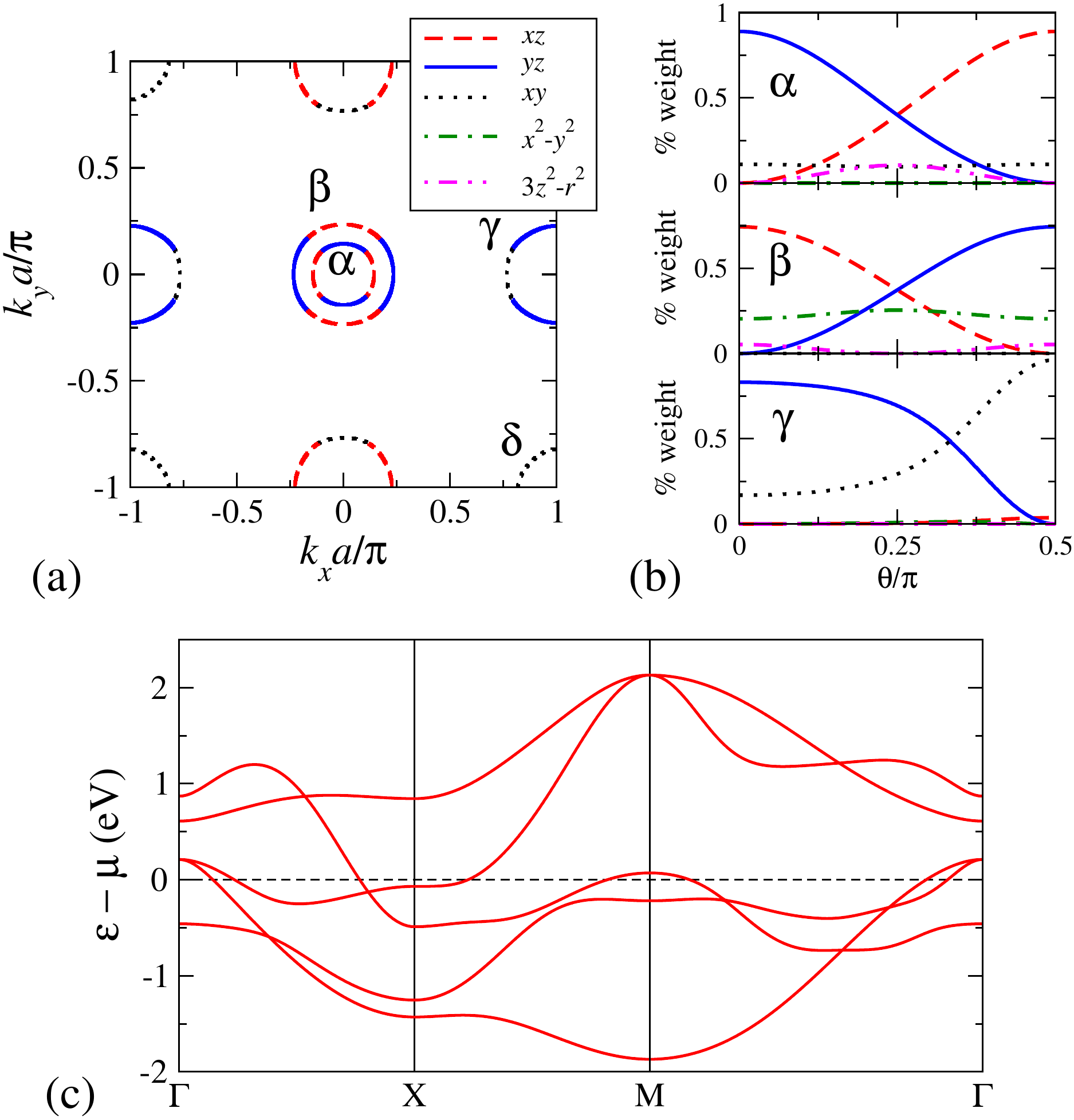}
  \end{center}
  \caption{\label{5bK_FS_BS} (a) The Fermi surface of the
    five-orbital model of Kuroki \emph{et al.}~\cite{Kuroki2008} showing the
    dominant orbital 
    contribution. (b) The 
    orbital weight around the $\alpha$, $\beta$ and $\gamma$ Fermi surfaces in
    (a); the orbital weight around the $\delta$ Fermi surface is not shown as
    it has almost complete $xy$ character. (c) The 
    band structure along high-symmetry directions.}  
\end{figure}

\begin{figure}
  \begin{center}
  \includegraphics[clip,width=\columnwidth]{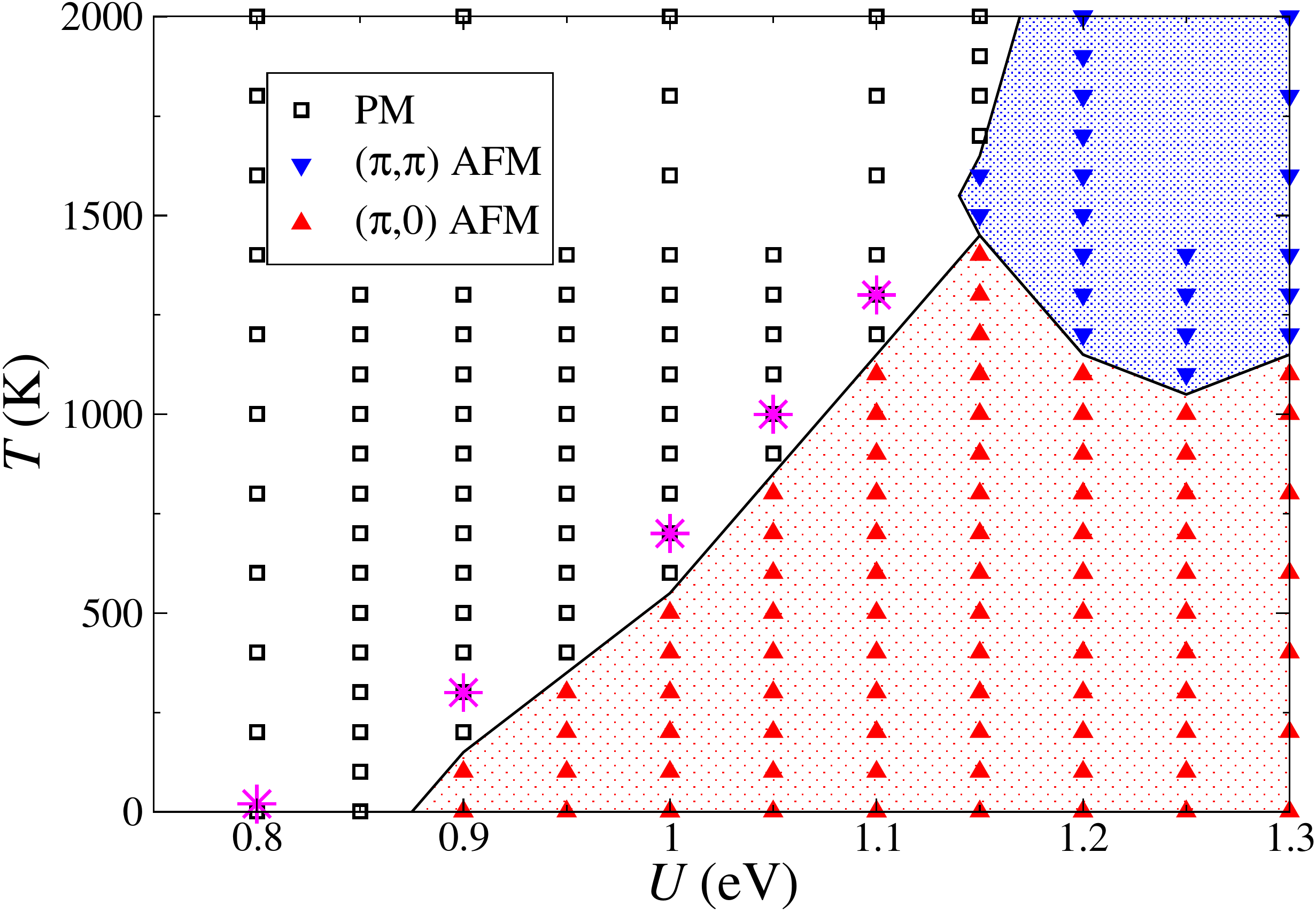}
  \end{center}
  \caption{\label{mf_5bK} The $T$-$U$ mean-field phase diagram of
    the five-orbital model of Kuroki \emph{et al.}.~\cite{Kuroki2008} The
    lightly-shaded region 
    approximately indicates the 
    extent of the $(\pi,0)$ AFM phase, while the darkly-shaded region gives
    the extent of the $(\pi,\pi)$ AFM phase. We show the static susceptibility
    at the points indicated by the star symbols in~\fig{chi_5bK}.}  
\end{figure}

This model was originally constructed in an orbital basis where
the $x$ and $y$ axes are rotated by 45$^{\circ}$ to the Fe lattice. In the
other models considered here, however, the Fe lattice defines the coordinate
axes for the orbital basis. 
To allow direct comparison with these other models, we therefore 
show the Fermi surface of the five-orbital model of Kuroki \emph{et al.} in
the usual basis in~\fig{5bK_FS_BS}(a). Although the Fermi surface has
the same basic form and orbital composition as the three-orbital model, the
electron pockets are almost 
circular, and there is an additional $xy$-dominated hole pocket
at the M point. Furthermore, from the band
structure~\fig{5bK_FS_BS}(c) we see that
the hole pockets at the $\Gamma$
point have similar Fermi velocities to the electron pockets, which
themselves 
have much higher Fermi velocity along the $\Gamma-$X line than in the X$-$M
direction. The latter accounts for the anisotropic spin-wave
dispersion found in~\Ref{Kaneshita2010}. 

\begin{figure*}
  \begin{center}
  \includegraphics[clip,width=2\columnwidth]{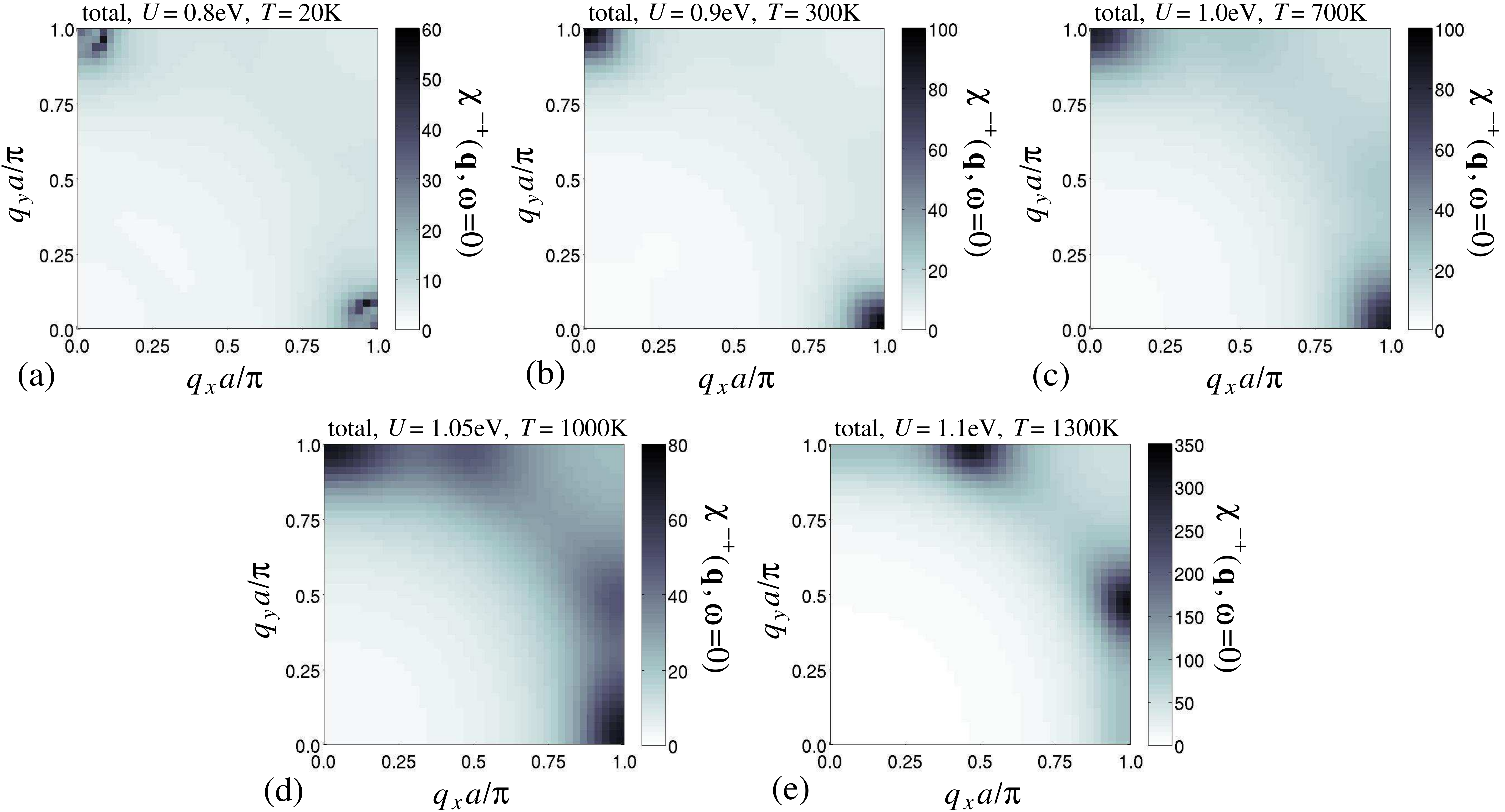}
  \end{center}
  \caption{\label{chi_5bK} The total static transverse spin
    susceptibility of the five-orbital model of Kuroki \emph{et
      al.}~\cite{Kuroki2008} at 
    (a) $U=0.8$\,eV and $T=20$\,K, (b) $U=0.9$\,eV and $T=300$\,K, (c) $U=1.0$\,eV and
    $T=700$\,K, (d) $U=1.05$\,eV and $T=1000$\,K, and (e) $U=1.1$\,eV and $T=1300$\,K. }  
\end{figure*}

In constructing our mean-field phase diagram~\fig{mf_5bK} we extended
the ansatz~\eq{eq:mfansatz} to also compare the free energy of the ordered
states within the rotated orbital basis used by Kuroki \emph{et al.} to
those in the usual orbital basis. We 
find that the mean-field solution in the usual orbital basis always has the
same [PM and $(\pi,\pi)$ AFM] or slightly lower [$(\pi,0)$ AFM] free energy
than the 
mean-field solution in the rotated basis. Although our results are not
directly comparable with those of Kaneshita \emph{et al.},~\cite{Kaneshita2009}
who used the rotated orbital basis and allowed for inter-orbital mean
fields, our phase diagram is nevertheless in good qualitative agreement:~\cite{Kaneshita2009,Kariyado2009} for $U_{c}>0.85$\,eV we find a
$(\pi,0)$ AFM state whose critical temperature rapidly grows with increasing
$U$. At $U\gtrsim1.15$\,eV and $T>1000$\,K, however, the
$(\pi,\pi)$ AFM state is stable.  

In~\fig{chi_5bK} we
present the results for the static transverse susceptibility at the points
marked in~\fig{mf_5bK}. For $U\lesssim1$\,eV, the results are rather similar to
those already seen in the two- and four-orbital models. In the low-temperature
PM state [\fig{chi_5bK}(a)], the susceptibility takes a maximum in an arc of
radius $\approx0.05\pi$ about $(\pi,0)$; at slightly
higher $U=0.85$\,eV we find evidence of an incommensurate AFM state with
ordering vector $(0.925\pi,0.025\pi)$. For the
points lying directly above the mean-field $(\pi,0)$ AFM state, however, the
susceptibility takes a maximum exactly at ${\bf q}=(\pi,0)$. This
peak is nevertheless much broader than in the two- or four-orbital models,
indicating a larger distribution of nesting vectors.

\begin{figure*}
  \begin{center}
  \includegraphics[clip,width=2\columnwidth]{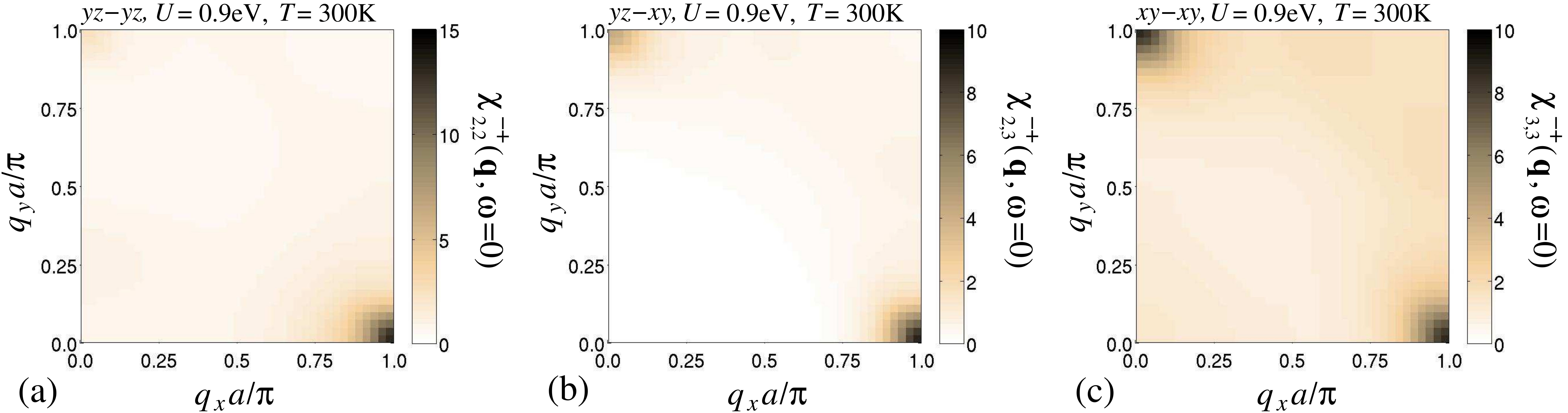}
  \end{center}
  \caption{\label{chi_5bK_dom} The dominant contributions to the
    $(\pi,0)$-peak at $U=0.9$\,eV and $T=300$\,K in the five-orbital model of
    Kuroki \emph{et al.}:~\cite{Kuroki2008} the (a) $yz$-$yz$, (b) $yz$-$xy$,
    and (c) $xy$-$xy$ susceptibilities.}  
\end{figure*}

The dominant contributions to the peak at ${\bf q}=(\pi,0)$ in the $U=0.9$\,eV,
$T=300$\,K static susceptibility [\fig{chi_5bK}(b)] are shown
in~\fig{chi_5bK_dom}. The most important terms are the $yz$-$yz$
[\fig{chi_5bK_dom}(a)], $yz$-$xy$, [\fig{chi_5bK_dom}(b)]
and $xy$-$xy$  [\fig{chi_5bK_dom}(c)] susceptibilities, which together constitute $\approx50$\% of the
peak height in the total susceptibility; for comparison, in the four-orbital
model the dominant susceptibilities shown in~\fig{chi_4b}(d-f) are responsible
for $\approx90$\% of the $(\pi,0)$ peak. The dominant terms in Kuroki \emph{et
  al.}'s five-orbital model are consistent with nesting of the outer hole
pocket with the electron pockets being the main driver of the magnetism, 
but with the nesting of the hole pocket at the M point with the electron
pockets also playing a substantial role. This is supported by the
reconstructed Fermi surface shown in~\Ref{Kaneshita2009} for the low-moment
case. 

As $U$ is increased to $1.05$\,eV [\fig{chi_5bK}(d)] the nesting picture of the
magnetism begins to break down, with the appearance of a definite (but still
subdominant) peak at ${\bf q} = (\pi,0.5\pi)$. This
feature completely dominates the response at $U=1.1$\,eV, indicating
that the 
mean-field phase diagram is unreliable for higher coupling
strengths. Peaks at these wavevectors have previously been
observed,~\cite{Arita2009,Kuroki2009} where they were explained as due to
scattering between the electron Fermi pockets. This appears to be inconsistent
with the observation that these peaks are most pronounced at high temperatures
$\gtrsim 1000$\,K, which instead suggests that band structure features
\emph{away} from the Fermi surface are responsible. In particular, we note
that there is a region of flat bands at $-0.2$\,eV near the M point, which is
predominantly derived from the almost-filled $3z^2-r^2$
orbital.~\cite{Ikeda2008,Arita2009,Ikeda2010} These states are shifted to
higher energies by the Hartree terms, and hence might play a major role in the
magnetic response at high temperatures. This will be discussed in detail
in~\Sec{subsec:HS}. 

\subsection{The five orbital model of Graser \emph{et
    al.}} \label{subsec:Graser}

\begin{figure}
  \begin{center}
  \includegraphics[clip,width=\columnwidth]{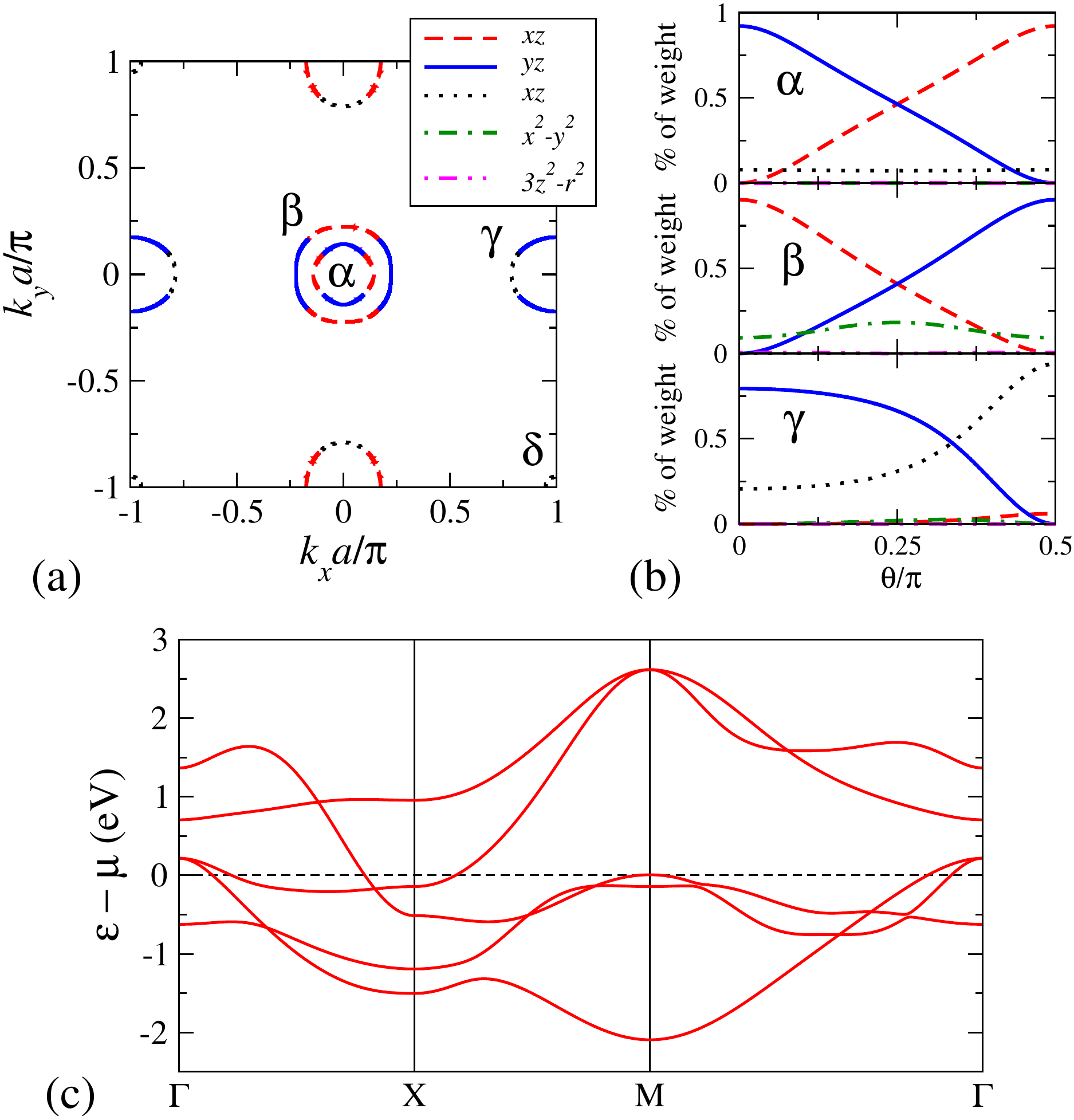}
  \end{center}
  \caption{\label{5bG_FS_BS} (a) The Fermi surface of the
    five-orbital model of Graser \emph{et al.}~\cite{Graser2009} showing the
    dominant orbital 
    contribution. (b) The 
    orbital weight around the $\alpha$, $\beta$ and $\gamma$ Fermi surfaces in
    (a); the orbital weight around the $\delta$ Fermi surface is not shown as
    it has almost complete $xy$ character. (c) The 
    band structure along high-symmetry directions.}  
\end{figure}

The five-orbital model of Graser \emph{et al.}~\cite{Graser2009} is a
frequently studied alternative to Kuroki \emph{et al.}'s model.
Graser \emph{et al.}'s model was also constructed by fitting to
\emph{ab initio} results; 
as can be seen from the electronic structure shown in~\fig{5bG_FS_BS}, both
five-orbital models share many similarities. The key
differences in the Fermi surface are that the electron pockets in this model
are clearly elliptical, while the hole pocket at the M point is almost
vanishingly small.~\cite{Grasernote} The former is in much better agreement
with current 
ARPES and \emph{ab initio} results; the latter refinement is also in 
reasonable agreement with 
\emph{ab initio} calculations, but the number of hole pockets remains somewhat
unclear in 
ARPES.~\cite{Shimojima2010,Zhang2010,Boeri2008,Eschrig2009,Yi2009a}
The flat band region near the M point is also closer to the Fermi surface than 
in Kuroki \emph{et al.}'s model; as such, we might expect it to play a larger
role in the magnetic response.

\begin{figure}
  \begin{center}
  \includegraphics[clip,width=\columnwidth]{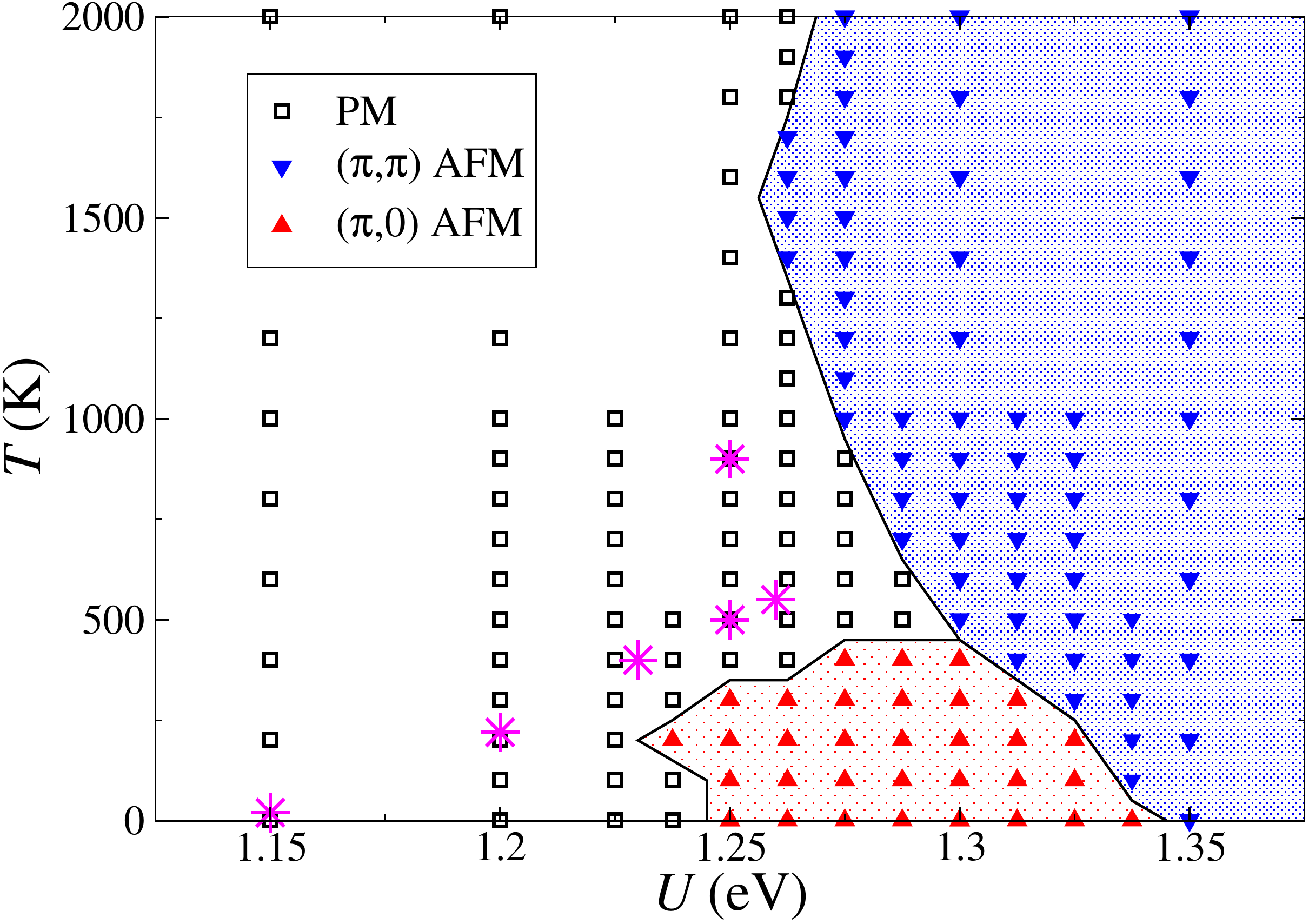}
  \end{center}
  \caption{\label{mf_5bG} The $T$-$U$ mean-field phase diagram of
    the five-orbital model of Graser \emph{et al.}.~\cite{Graser2009} The
    lightly-shaded region 
    approximately indicates the 
    extent of the $(\pi,0)$ AFM phase, while the darkly-shaded region gives
    the extent of the $(\pi,\pi)$ AFM phase. We show the static susceptibility
    at the points indicated by the star symbols in~\fig{chi_5bG}.}  
\end{figure}

The magnetic properties of Graser \emph{et al.}'s five-orbital model has
only recently attracted attention.
In~\Ref{Daghofer2010b} it was shown that the $(\pi,0)$ AFM state was
realized at $T=0$\,K for $U \gtrsim 1.23$\,eV, but the authors did not consider
competition with other magnetic states. The existence of a $(\pi,0)$ AFM
ground state with realistic ordered moment and Fermi surface was confirmed
in~\Ref{Luo2010} for a rather narrow range of $U$. For much larger
  values of $U$ it has been shown that Hartree-Fock and Gutzwiller
theories give quite different results.~\cite{Schickling2010}
In~\fig{mf_5bG} we show the phase
diagram in the $T$-$U$ plane: a $(\pi,0)$ AFM state with low critical
temperature $<500$\,K is realized for $1.25$\,eV$\lesssim U \lesssim$1.35\,eV, but the phase diagram
is clearly dominated by a $(\pi,\pi)$ AFM state with high critical temperature
$>2000$\,K. For most of the $U$ range where the $(\pi,0)$ AFM state is stable at
$T=0$\,K, we find that the $(\pi,\pi)$ AFM state is in fact realized at higher
$T$, with the PM phase separating them at lower values of $U$.

\begin{figure*}
  \begin{center}
  \includegraphics[clip,width=2\columnwidth]{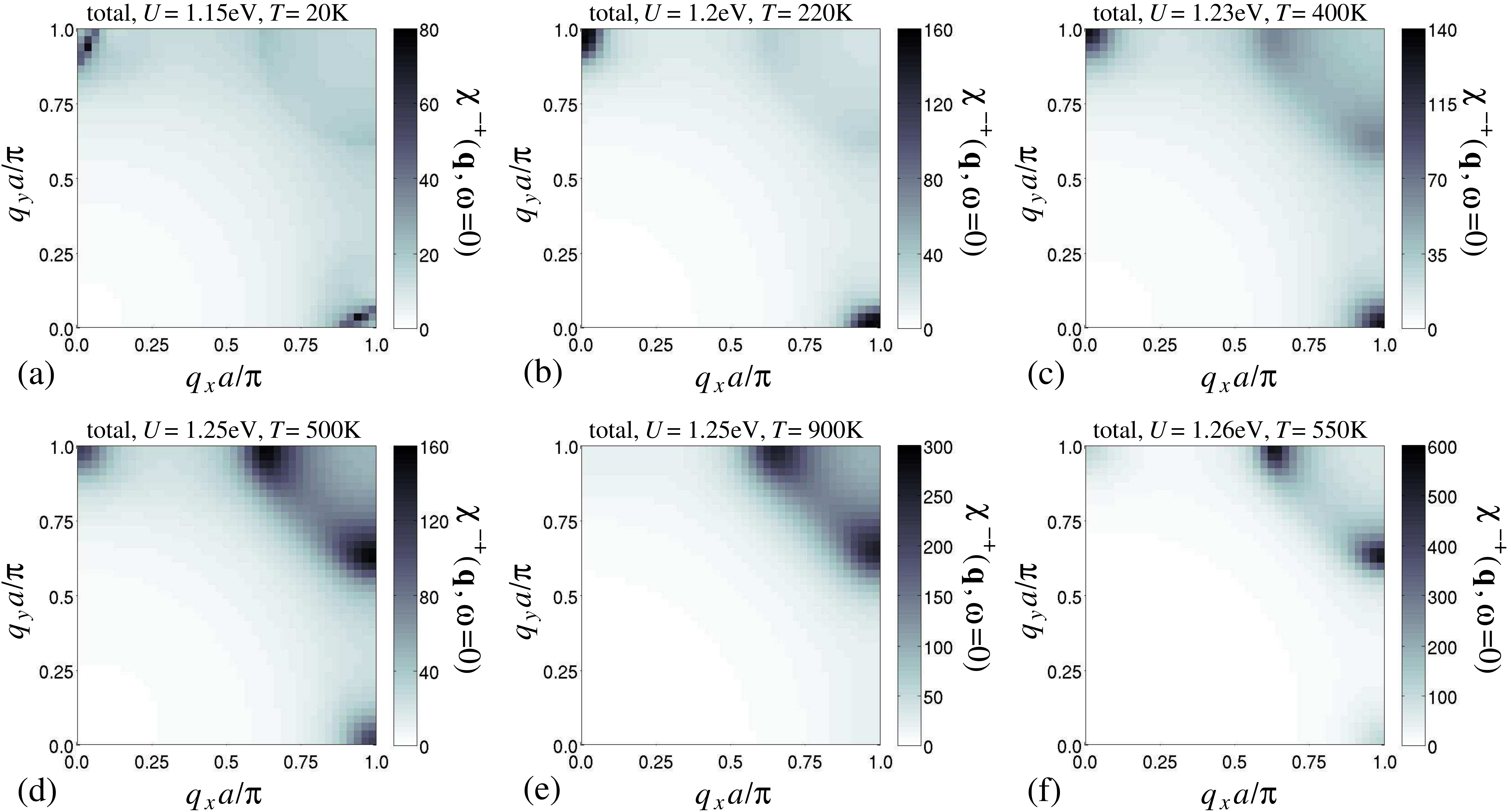}
  \end{center}
  \caption{\label{chi_5bG} The total static transverse spin
    susceptibility of the five-orbital model of Graser \emph{et
      al.}~\cite{Graser2009} at 
    (a) $U=1.15$\,eV and $T=20$\,K, (b) $U=1.2$\,eV and $T=220$\,K, (c) $U=1.23$\,eV and
    $T=400$\,K, (d) $U=1.25$\,eV and $T=500$\,K, (e) $U=1.25$\,eV and $T=900$\,K, and
    (f) $U=1.26$\,eV and $T=550$\,K. }  
\end{figure*}

The results for the static susceptibility [\fig{chi_5bG}] indicate that the
mean-field phase diagram is only partially correct. The evolution of
$\chi^{-+}({\bf q},\omega=0)$ up to the edge of the $(\pi,0)$ AFM phase is
similar to that in Kuroki \emph{et al.}'s model: as $U$ is increased, a
definite tendency to weakly-incommensurate order [with a peak at ${\bf
    q}\approx(0.95\pi,0.025\pi)$] 
gives way to broad peaks at ${\bf q}=(\pi,0)$. We have confirmed an
instability towards incommensurate order below at least $100$\,K at
$U=1.2$\,eV. We also note a ring-like feature centred at ${\bf
  q}=(\pi,\pi)$ in~\fig{chi_5bG}(a); by $U=1.23$\,eV this has
apparently evolved into a broad peak at ${\bf 
  q}=(\pi,0.6\pi)$. Although this is similar to the additional peaks seen 
in the high-temperature results for Kuroki \emph{et al.}'s model, they appear
here at much lower temperatures. 

\begin{figure*}
  \begin{center}
  \includegraphics[clip,width=2\columnwidth]{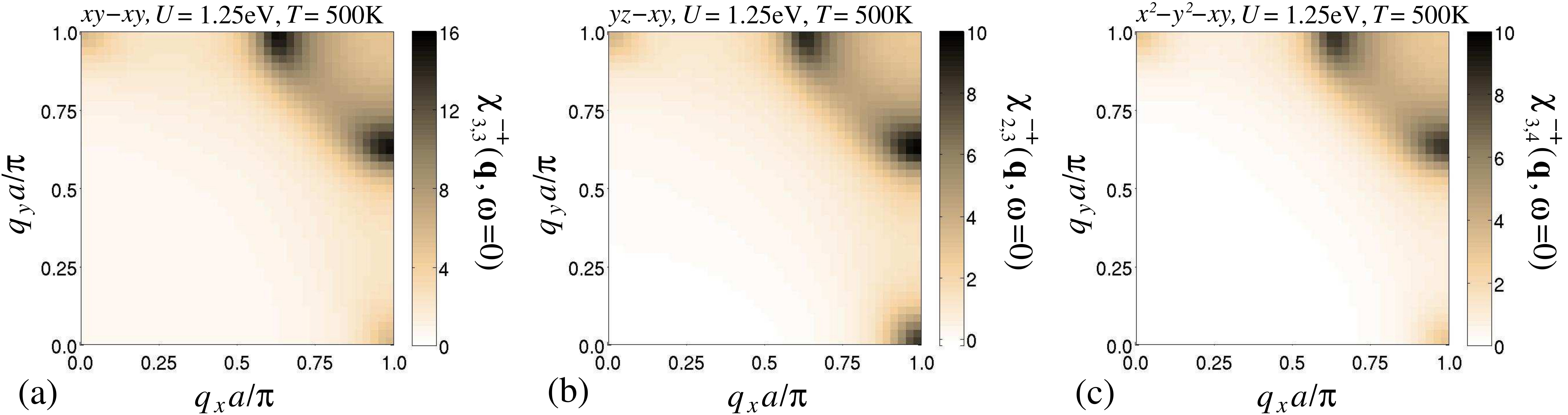}
  \end{center}
  \caption{\label{chi_5bG_dom} The dominant contributions to the
    static transverse spin
    susceptibility at $U=1.25$\,eV, $T=500$\,K in the five-orbital model of Graser
    \emph{et al.}:~\cite{Graser2009} 
    (a) $xy$-$xy$, (b) $yz$-$xy$, and (c) $x^2-y^2$-$xy$ susceptibilities.}  
\end{figure*}

The peak at $(\pi,0.6\pi)$ displays a peculiar
temperature-dependence. To see this, in~\fig{chi_5bG}(d) and (e) we
show the 
static susceptibility at $U=1.25$\,eV for $T=500$\,K and $900$\,K,
respectively. In the former case, the peaks at both $(\pi,0)$ and
$(\pi,0.6\pi)$ are distinctly visible. We have
verified that as the temperature is lowered to $\sim450$\,K the $(\pi,0)$ peak
diverges while the $(\pi,0.6\pi)$ peak is somewhat suppressed, and so the
$(\pi,0)$ AFM state is realized at low temperatures in agreement with
the mean-field phase diagram. As the temperature is raised to
$T=900$\,K, however, the peak at $(\pi,0)$ disappears while the $(\pi,0.6\pi)$
peak almost doubles in height, 
revealing that the system is close to an instability towards a
$(\pi,0.6\pi)$ 
AFM state. This also suggests that the mechanism responsible for the
$(\pi,0.6\pi)$ peak is not due to scattering between different Fermi 
pockets, but rather due to the electronic structure away from the Fermi
surface. In~\fig{chi_5bG}(f) we show the static susceptibility at
slightly higher $U$ and $T$ compared to~\fig{chi_5bG}(d): the $(\pi,0.6\pi)$
peak completely dominates the 
response, and the system is again close to ordering at this
wavevector. Along  
with~\fig{chi_5bG}(e), this suggests that the $(\pi,\pi)$ AFM state
in~\fig{mf_5bG} should be replaced by a $(\pi,0.6\pi)$ AFM state; the 
region of stable $(\pi,0)$ AFM order will also likely shrink.

To investigate the origin of the $(\pi,0.6\pi)$ peak we have examined the
dominant contributions to $\chi^{-+}$. Focusing upon the
$U=1.25$\,eV, $T=500$\,K case, we find that the
susceptibilities involving the $xy$-orbital are most important, contributing
$\approx 50$\% of the peak height; the largest contributions come from
the $xy$-$xy$ term [\fig{chi_5bG_dom}(a)], the $yz$-$xy$ term
[\fig{chi_5bG_dom}(b)], the $xz$-$xy$ term [\fig{chi_5bG_dom}(b) with
  $q_x\leftrightarrow q_y$], and the $(x^2-y^2)$-$xy$ term
[\fig{chi_5bG_dom}(c)]. The dominant role of the $xy$ orbital is consistent
with the results of~\Ref{Arita2009}
and~\Ref{Kuroki2009} for similar orbital models, but it is difficult
to reconcile the  
temperature-dependence observed here with the author's interpretation 
that the $(\pi,0.6\pi)$ peak originates from
scattering between the electron pockets. An alternative possibility is
scattering between the hole pockets at the $\Gamma$ point and the $xy$-derived
hole pocket at the M point, although the required wavevector is slightly too
large. We will discuss this  matter further in~\Sec{subsec:HS}.

\section{Discussion} \label{sec:discussion}

To summarize our main results, we found that the
two-orbital,~\cite{Raghu2008} four-orbital,~\cite{Yu2009} and 
Kuroki \emph{et 
  al.}'s five-orbital~\cite{Kuroki2008} models display a rather robust
$(\pi,0)$ AFM phase in 
their zero-doping phase diagrams. In both the four- and five-orbital models,
this AFM state competes with a strong-coupling AFM state with different
ordering vector.
In contrast, the three-orbital~\cite{Daghofer2010a} and Graser \emph{et al.}'s
five-orbital~\cite{Graser2009} models 
show at most a weak tendency towards $(\pi,0)$ AFM order: in the
former it is 
likely not present anywhere in the considered phase diagram, while in
the latter it is out-competed over much of the phase diagram by a much more
stable $(\pi,0.6\pi)$ AFM state. 

Our results support the scenario of the $(\pi,0)$ AFM state
originating from nesting of the electron and hole Fermi pockets. This is a
crucial ingredient for the magnetism: as discussed below
in~\Sec{subsec:nesting}, the stability of the $(\pi,0)$ AFM state is
  positively correlated with the degree of the nesting. The similarity of the
electronic  structure of the 
two five-orbital models, yet the very large differences in their magnetic
phase diagrams, is also of special note. In particular, this sheds light on the
origin of the $(\pi,0.5\pi)$-$(\pi,0.6\pi)$ AFM state realized in these
models at strong 
coupling or high temperatures. In~\Sec{subsec:HS} we argue that subtle
differences in the states derived from the $3z^2-r^2$ orbital play a
significant role in stabilizing this state. Finally,
in~\Sec{subsec:implications} we discuss the 
implications of our results for models of the magnetic order in the iron
pnictides. 

\subsection{Nesting} \label{subsec:nesting}

\begin{figure*}
  \begin{center}
  \includegraphics[clip,width=2\columnwidth]{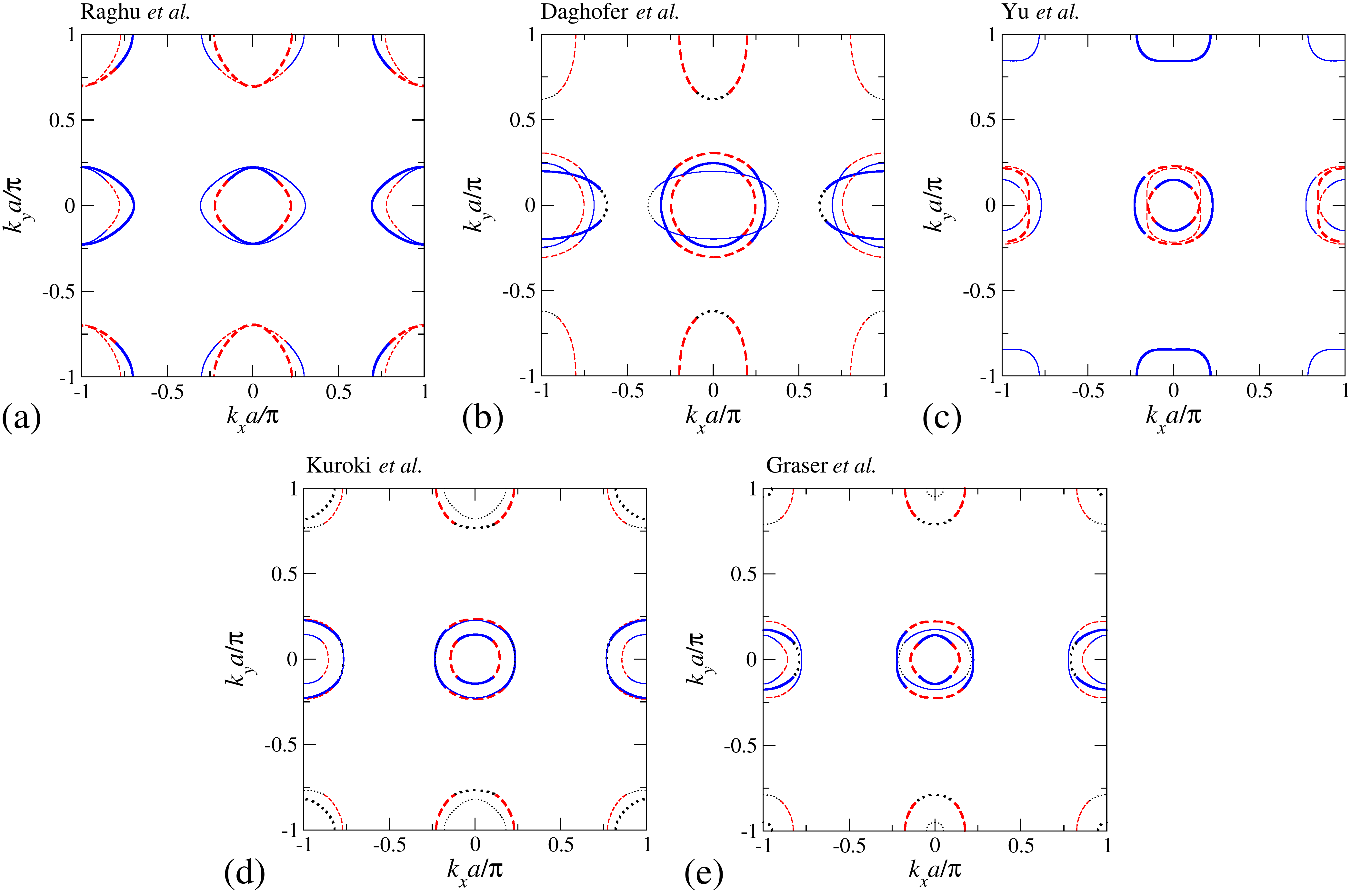}
  \end{center}
  \caption{\label{nesting} Nesting properties of the orbital
    models. We show the Fermi surface (thick lines) and the Fermi surface
    translated by $(\pi,0)$ (thin lines) of (a) the two-orbital model of Raghu
    \emph{et al.},~\cite{Raghu2008} (b) the
  three-orbital model of Daghofer \emph{et al.},~\cite{Daghofer2010a} (c) the
  four-orbital model of Yu \emph{et al.},~\cite{Yu2009} (d) the five-orbital
  model  
  of Kuroki \emph{et al.},~\cite{Kuroki2008} and (e) the five-orbital model of
  Graser \emph{et al.}.~\cite{Graser2009}}  
\end{figure*}

The $(\pi,0)$ AFM order was found to be most robust in models where the  
quality of the nesting between the electron and hole pockets is high. For
perfect nesting the dispersion $\epsilon_{\bf k}$ satisfies 
\beq
\epsilon_{\bf k}=-\epsilon_{{\bf k}+{\bf Q}}\, ,  \label{eq:nesting}
\eeq
for some nesting vector ${\bf Q}$.
In such a scenario, not only do the Fermi surfaces exactly
overlap upon performing the shift by ${\bf Q}$, but also the holes and
electrons have the same effective mass. In realistic models such as those
considered here the condition~\eq{eq:nesting} is only approximately
fulfilled. For a nesting picture to still be relevant, we require that some
segments of the electron and hole
Fermi surfaces should overlap, and that the effective mass difference between
them should not be too great. We note that it is sometimes the case that while
the condition~\eq{eq:nesting} is satisfied approximately for the entire Fermi
surface for the vector ${\bf Q}$, a better match between some Fermi
surface segments (but worse for others) may be achieved for a slightly
different vector $\widetilde{\bf
  Q}\approx{\bf Q}$. At weak coupling strengths, where the effective staggered
magnetic potential is 
small, more of the Fermi surface may be gapped at ordering vector
$\widetilde{\bf Q}$ than at ${\bf Q}$.

In~\fig{nesting}, we show the Fermi surface of each
of the models superimposed with the  Fermi surface translated by ${\bf
  Q}=(\pi,0)$. 
It is immediately clear that the three-orbital model has the worst nesting
properties: not only is there very poor overlap between the hole Fermi surface
and the translated electron Fermi surface, but also the electron pocket
has much lower effective mass than either of the hole pockets. It is therefore
not surprising that this is the only model that fails to show the required
$(\pi,0)$ AFM order.

Although both electron pockets participate in the $(\pi,0)$ AFM state in the
two-orbital model, the nesting here is nevertheless rather similar to that in
the four-orbital model. In these two models, a segment near the minor axis of
the shifted elliptical electron pocket overlaps with the small hole pocket,
while a segment near the major axis overlaps with the large hole pocket
[see~\fig{nesting}(a) and (c)]. There is an excellent match between the
dominant orbital character of the original and the shifted Fermi surfaces,
which should enhance the AFM fluctuations as the intra-orbital Coulomb
repulsion is much larger than the inter-orbital Hund's rule coupling. 

The nesting of the electron pockets with the outer hole pocket at the
$\Gamma$ point in Kuroki 
\emph{et al.}'s five-orbital model is the best considered here, with almost
complete overlap with the shifted Fermi surfaces. The orbital-resolved
susceptibilities suggests that this is most important to the
magnetic instability, although the poorer nesting of the electron pockets with
the hole pocket at the M point is likely
responsible 
for broadening the peak at $(\pi,0)$. Unlike the two- and four-orbital 
models, therefore, there is a relatively poor match of 
the dominant orbital character of the nested Fermi surfaces. This is reflected
in the susceptibilities: $\sum_{\nu}\chi^{-+}_{\nu,\nu}$ contributes
$\approx30$\% of the height of the $(\pi,0)$ peak in 
Kuroki \emph{et al.}'s model, while in the four-orbital model the
susceptibility $\chi^{-+}_{1,1}$ is alone responsible for $\approx50$\% of the
peak height. In contrast, despite the near-identical orbital
content of the Fermi surfaces, 
the instability towards $(\pi,0)$ AFM order in the five-orbital model of
Graser \emph{et al.} is much weaker. This can be explained by the absence of
direct Fermi-surface overlap between the electron and the hole pockets at the
$\Gamma$ point, and also the much
weaker nesting between electron pockets and the small hole pocket at the M
point. The stark differences between the two five-orbital models highlights
the sensitivity of the 
$(\pi,0)$ AFM state to subtle details of the electronic structure.

Our results suggest that a qualitatively-correct understanding of the AFM order
in the pnictides can be developed without reference to the orbital
structure, i.e. they support the excitonic theories of the magnetism. This is
not unexpected, as the $(\pi,0)$ AFM order is realized at
relatively weak coupling strengths, where the excitonic model arises naturally
as a 
low-energy effective theory.~\cite{Chubukov2008,Cvetkovic2009} The rather
strong competition with other AFM states, however, is not anticipated by such
excitonic models. 

\subsection{Hartree shifts} \label{subsec:HS}

\begin{figure*}
  \begin{center}
  \includegraphics[clip,width=2\columnwidth]{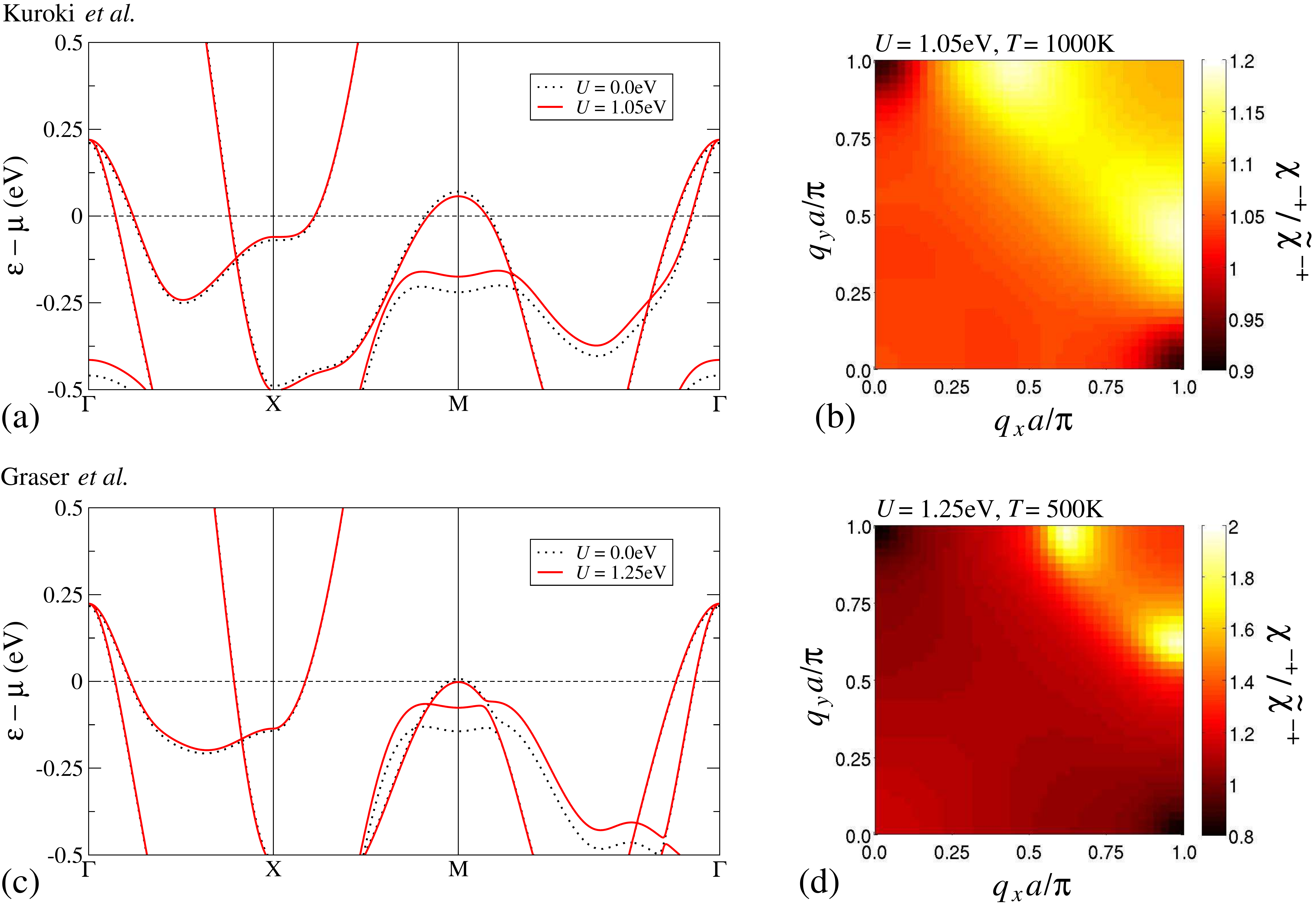}
  \end{center}
  \caption{\label{HS} Kuroki \emph{et al.}'s five-orbital
    model:~\cite{Kuroki2008} 
    (a) band structure along high-symmetry
    directions with and without the Hartree shift at $U=1.05$\,eV, $T=0$\,K; (b)
    ratio of the static transverse susceptibility with ($\chi^{-+}$) and without
    ($\widetilde{\chi}^{-+}$) the Hartree shift included for $U=1.05$\,eV,
    $T=1000$\,K. Graser \emph{et al.}'s five-orbital model:~\cite{Graser2009}
    (c) band structure 
    along high-symmetry 
    directions with and without the Hartree shift at $U=1.25$\,eV, $T=0$\,K; (d)
    ratio of the static transverse susceptibility with ($\chi^{-+}$) and without
    ($\widetilde{\chi}^{-+}$) the Hartree shift included for $U=1.25$\,eV,
    $T=500$\,K.} 
\end{figure*}

The inclusion of Hartree shifts in our mean-field decoupling scheme
requires justification: models where the tight-binding band structure is
constructed by fitting to
\emph{ab initio} results should already account for the Hartree shifts,
so it is then only necessary to include the staggered magnetic
potentials in~\eq{eq:mfham}. It is nevertheless likely that such methods
underestimate the effect of the correlations; indeed, 
comparison with ARPES reveals that the bands obtained within the local density
approximation are subject to significant
energy shifts.~\cite{Yi2009b} Including the Hartree shifts is the simplest
way to test the robustness of the AFM order towards such modifications of
the band structure. In~\Ref{Kuroki2009} it was reported that the
Hartree shifts can lead to dramatic changes in the Fermi surface of Kuroki
\emph{et al.}'s five-orbital model. Ikeda and co-workers similarly found
unphysical renormalization of the Fermi surface by the $\omega=0$
self-energy corrections calculated within the fluctuation-exchange
approximation.~\cite{Ikeda2008,Arita2009,Ikeda2010}

In order to evaluate the effect of the Hartree shifts, we have calculated the
renormalized band structures for each model in a fictitious $T=0$\,K PM
state at 
a coupling strength where an AFM mean-field state is realized. This is a good
estimate of the band structure in the stable $T>0$\,K PM state as there is only
rather small temperature-dependence of the orbital occupancies. The
  degeneracy 
of the $xz$ and $yz$ orbitals in the two-orbital model means that there is no
change in the band structure; for the three- and four-orbital models we find
only very slight changes in the band structure far from the Fermi surface
($\gtrsim0.5$\,eV and $1$\,eV, respectively). As such, the inclusion of the
Hartree shifts should be irrelevant to the weak-coupling magnetic
response of these models.  

For the five-orbital models we again find very minor changes in the Fermi
surface, limited to the shrinking (Kuroki \emph{et al.}) or removal (Graser
\emph{et al.}) of the $xy$-derived hole pocket at the M point, which is due to
the positive Hartree shift of the other orbitals relative 
to the (least-occupied) $xy$ orbital. As shown in~\fig{HS}(a) and (c),
however, the flat-band region near the M point and below the Fermi energy
undergoes an almost uniform shift to higher energies. These states are mostly
derived from the $3z^2-r^2$ orbital, which has the highest occupation and
hence has the largest Hartree
shift.~\cite{Ikeda2008,Graser2009,Arita2009,Ikeda2010} 
The shift to higher energies is particularly dramatic in Graser \emph{et
  al.}'s model, where the flat band region is shifted from $\approx-0.15$\,eV to
$\approx-0.07$\,eV; in Kuroki \emph{et al.}'s model the equivalent
feature is shifted from $\approx-0.21$\,eV to $\approx-0.17$\,eV. We
note that the Hartree shifts of the orbitals are in qualitative agreement
with the more advanced analysis of Ikeda
\emph{et al.},~\cite{Ikeda2008,Arita2009,Ikeda2010} although the change in
the band structure is less severe in our results.

The shift of the $(3z^2-r^2)$-derived states closer to the Fermi surface
suggests that they might play a
role in the high-$T$ and $U$ magnetic behaviour of the system, where
the $(\pi,0)$ AFM state competes with an enigmatic 
$(\pi,0.5\pi)$-$(\pi,0.6\pi)$ AFM 
state. 
This is contradicted by the result from~\Sec{subsec:Graser} that the peak at
$(\pi,0.6\pi)$ is mostly due to the $xy$ orbital, with the $3z^2-r^2$ orbital
making a relatively modest contribution. To isolate the effect of the
$3z^2-r^2$ orbital, we have therefore recalculated the
static susceptibility \emph{without} including the Hartree shifts,
$\widetilde{\chi}^{-+}({\bf q},\omega=0)$. In~\fig{HS}(b) and (d) we show the
ratio of the static susceptibility with the Hartree shift to that without for
the cases~\fig{chi_5bK}(d) ($U=1.05$\,eV, $T=1000$\,K, Kuroki \emph{et al.}'s
model) and~\fig{chi_5bG}(d) ($U=1.25$\,eV, $T=500$\,K, Graser \emph{et al.}'s
model). In both cases, we see that the peak at 
$(\pi,0.5\pi)$-$(\pi,0.6\pi)$ is
significantly enhanced by the inclusion of the Hartree shifts, especially so
for Graser \emph{et al.}'s model. Including the Hartree shifts also
leads to a 
reduction of the height of the peak at $(\pi,0)$, probably due to the
shrinking of the $xy$ pocket, highlighting the sensitive
dependence of the $(\pi,0)$ AFM state upon fine details of the nesting.  

As the rest of the band structure undergoes relatively little change upon the
inclusion of the Hartree shifts, the results~\fig{HS}(b) and (d) indicate that
the shift of the $3z^2-r^2$-derived flat band near the M point is
heavily involved in stabilizing the $(\pi,0.5\pi)$-$(\pi,0.6\pi)$ AFM
state. This is paradoxical: although the $3z^2-r^2$ orbitals are clearly
important to 
the appearance of the $(\pi,0.5\pi)$-$(\pi,0.6\pi)$ AFM state, other
orbitals dominate the magnetic response at the ordering vector. Furthermore,
the inclusion of the Hartree shifts enhances all the orbitally resolved
susceptibilities $\chi^{-+}_{\mu,\nu}$ by a similar factor. For example,
in the $U=1.25$\,eV, $T=500$\,K results for Graser
\emph{et al.}'s model, the ratio
$2.3\gtrsim\chi^{-+}_{\mu,\nu}/\widetilde{\chi}^{-+}_{\mu,\nu}\gtrsim1.8$ for
all $\mu$ and $\nu$, although it is maximal when the $3z^2-r^2$ orbital is
involved. This appears to indicate that the flat
band near 
the M point plays an indirect role in the magnetic ordering. We speculate that
this involves the strong peak in the density of states associated with this
band, which feeds into the response of the other bands due to the orbital
mixing. 

The sensitivity of the 
$(\pi,0.5\pi)$-$(\pi,0.6\pi)$ peak 
to the Hartree shift implies a strong dependency upon the choice of the ratio
$J/U$. For $J/U=0.25$ considered here, we have positive effective Hartree
shifts $\widetilde{\epsilon}_\nu = 0.25U(n_\nu - \min\{n_\mu\})$. Had we chosen
$J/U=0.2$, however, the Hartree shifts would vanish; for $J/U<0.2$ we find
\emph{negative} Hartree shifts, and the $3z^2-r^2$ orbital would
  therefore be shifted 
\emph{away} from the Fermi energy, likely suppressing the
$(\pi,0.5\pi)$-$(\pi,0.6\pi)$ 
AFM order. 

\subsection{Implications for models of the iron pnictides} \label{subsec:implications}

One of the aims of this work was to identify an ideal low-energy orbital model
of the pnictides, i.e. a model possessing Fermi surfaces in good quantitative
agreement with ARPES and \emph{ab initio} results, and also displaying robust
$(\pi,0)$ order. Unfortunately, none of the models studied here satisfy both
criteria. The models which display robust $(\pi,0)$ AFM order have rather
unrealistic elements in their electronic structure: the two-orbital model has a
$xz$/$yz$-derived hole pocket at the M point and also neglects the large $xy$
weight at the Fermi energy, in the four-orbital model the $xy$ weight is
also unrealistically small at the Fermi surface and the electron pockets do
not involve a band crossing, while the electron pockets 
in Kuroki \emph{et al}.'s five-orbital model are not elliptical. In contrast,
the three-orbital model qualitatively captures the predicted Fermi surface
topology and dominant orbital contributions, while Graser \emph{et al.}'s
five-orbital model is in good quantitative agreement, but these two models
show the weakest tendencies towards $(\pi,0)$ AFM order.

The sensitivity of the $(\pi,0)$ AFM state to small details of the band
structure raises an important question about the magnetic state in the
pnictides: if electron-hole nesting is indeed responsible for this state, why
is it realized in such a large 
range of compounds, each with subtly different Fermi
surfaces?~\cite{Ikeda2010} A possible answer is that the coupling to the
lattice 
degrees of freedom allows the system to fine-tune its Fermi surface. Indeed,
the magnetic properties of these compounds have been shown to be
strongly coupled 
to details of the lattice structure: the lattice 
constants and pnictogen height above the Fe plane obtained in \emph{ab initio}
calculations are
strongly affected by the iron moment,~\cite{Mazin2008,Yildirim2009} and varying 
the Fe-As-Fe bond angles can significantly alter the Fermi
surface.~\cite{Kuroki2009,Calderon2009a,Calderon2009b,Vildosola2008} It
is certainly suggestive that Kuroki \emph{et al.}'s model was fitted to
\emph{ab initio} results using experimental crystal parameters, whereas
Graser \emph{et al.}'s is based on structurally-relaxed
calculations.~\cite{Cao2008}
 
Moreover, a structural phase transition from a high-$T$ tetragonal to low-$T$
orthorhombic phase occurs either just above ($R$FeAsO compounds) or
at ($A$Fe$_2$As$_2$ compounds) the AFM transition
temperature.~\cite{1111coupling,122coupling,Huang2008,Lumsden2010} A scenario for a
joint magnetic-structural transition in 
the pnictides was proposed within the context of a simple excitonic model
in~\Ref{Barzykin2009}: the lattice distortion improves the nesting condition
sufficiently to allow the appearance of an AFM state, which in turn stabilizes
the distortion. This was verified for a more realistic 
four-band excitonic model in~\Ref{Brydon2009a}. Further work to clarify the
role of the orthorhombic distortion is clearly necessary. It would be
particularly interesting to examine this question within an orbital model, as 
the directional wave functions of the orbitals naturally 
couple to the lattice. The possibility that spontaneous orbital
ordering drives the structural transition is of particular
interest.~\cite{Lv2009,Chen2010}  
Of course, another possible explanation for the stability of the
$(\pi,0)$ AFM order involves strong-coupling physics beyond the nesting
scenario.~\cite{Zhou2010,Schickling2010} We note that recent calculations
using dynamical mean-field theory 
suggest that the correlation 
strength in these materials has been underestimated.~\cite{Hansmann2010}
Further research is clearly needed to understand the complex magnetic
physics of the pnictides.

\section{Summary} \label{sec:summary}

In this work we have presented a systematic weak-coupling investigation
into the 
appearance of the expected $(\pi,0)$ AFM 
order in five different orbital models of the iron pnictide parent
compounds. 
The mean-field phase diagrams of each model were first determined,
and then the static spin susceptibility was calculated within RPA 
in the PM state close to the boundaries of the magnetic phases. The highest
peaks in the spin susceptibility reveal the actual ordering vector of the
nearby mean-field state. The origin of these peaks was studied by
examining the dominant orbitally resolved contributions to the total spin
susceptibility. This procedure allowed an unbiased assessment of the
magnetic-ordering properties of each model. 

The studied models display an unexpectedly rich range of magnetic
behaviour. Only four of the models were found to have an instability to an AFM
state with  
the required ordering vector, although the robustness of this phase varies
greatly between them. In each of these models we also uncovered evidence of a
weakly-incommensurate AFM phase with ordering vector close to $(\pi,0)$ at
coupling strengths below the minimum required for a stable $(\pi,0)$
state. The $(\pi,0)$ AFM state originates from the  
nesting of the electron and hole Fermi pockets. The quality of the
nesting appears to be of primary relevance for the stability of the $(\pi,0)$
AFM 
order, while the matching of the orbital character of the nested Fermi surfaces
is of lesser importance. In the most realistic five-orbital models
the $(\pi,0)$ AFM 
state was found to compete with a strong-coupling incommensurate magnetic
state. We have argued that the origin of this phase
involves electronic structure below the Fermi energy. The implications of our
results for orbital models of the pnictides was discussed, and 
we propose that the apparent sensitivity
of the $(\pi,0)$ AFM order to fine details of the low-energy electronic
structure indicates that 
lattice degrees of freedom should be included in a theoretical 
description of the magnetism of the pnictides.

\begin{acknowledgments}
The authors thank I. Eremin, J. Knolle, K Kuroki, and P. Thalmeier for useful
discussions. PMRB and CT acknowledge funding from the DFG priority program
1458 and MD from the Emmy-Noether program of the DFG.
\end{acknowledgments}


\begin{thebibliography}{99}

\bibitem{Kamihara2008}Y. Kamihara, T. Watanabe, M. Hirano, and H. Hosono,
  J. Am. Chem. Soc. {\bf 130}, 3296 (2008).

\bibitem{Rotter2008}M. Rotter, M. Tegel, and D. Johrendt,
  Phys. Rev. Lett. {\bf 101}, 107006 (2008).

\bibitem{1111coupling}J. Zhao, Q. Huang, C. de la Cruz, S. Li,
  J. W. Lynn, Y. Chen, M. A. Green, G. F. Chen, G. Li, Z. Li, J. L. Luo,
  N. L. Wang, P. Dai, Nature Materials {\bf{7}}, 953
  (2008); J. Zhao, Q. Huang, C. de la Cruz, J. W. Lynn, M. D. Lumsden,
  Z. A. Ren, J. Yang, X. Shen, X. Dong, Z. Zhao, P. Dai, Phys. Rev. B
  {\bf{78}}, 132504 (2008). 

\bibitem{122coupling}A. Jesche, N. Caroca-Canales, H. Rosner, H. Borrmann,
  A. Ormeci, D. Kasinathan, K. Kaneko, H. H. Klauss, H. Luetkens, R. Khasanov,
  A. Amato, A. Hoser, C. Krellner, and C. Geibel, Phys. Rev. B {\bf{78}},
  180504(R) (2008).

\bibitem{Huang2008}Q. Huang, Y. Qiu, W. Bao, M. A. Green, J. W. Lynn,
  Y. C. Gasparovic, T. Wu, G. Wu, and X. H. Chen, Phys. Rev. Lett. {\bf 101},
  257003 (2008).

\bibitem{Lee2006}P. A. Lee, N. Nagaosa, and X.-G. Wen, Rev. Mod. Phys. {\bf
  78}, 17 (2006).

\bibitem{Lumsden2010}M.D. Lumsden and A.D. Christianson, J. Phys.:
  Condens. Matter {\bf 22}, 203203 (2010).

\bibitem{Johnston2010}D. C. Johnston, Adv. Phys. {\bf 59}, 803 (2010).

\bibitem{delaCruz2008}C. de la Cruz, Q. Huang, J. W. Lynn, J. Li, W. Ratcliff
  II, J. L. Zarestky, H. A. Mook, G. F. Chen, J. L. Luo, N. L. Wang, and
  P. Dai, Nature {\bf 453}, 899 (2008).

\bibitem{Pratt2011}D. K. Pratt, M. G. Kim, A. Kreyssig, Y. B. Lee,
  G. S. Tucker, A. Thaler, W. Tian, J. L. Zarestky, S. L. Bud'ko,
  P. C. Canfield, B. N. Harmon, A. I. Goldman, and R. J. McQueeney,
  arXiv:1104.0717

\bibitem{magneto}S. E. Sebastian, J. Gillett, N. Harrison, P. H. C. Lau,
  C. H. Mielke, and G. G. Lonzarich, J. Phys:
  Condens. Matter {\bf{20}}, 422203 (2008); J. G. Analytis, R.
  D. McDonald, J.-H. Chu, S. C. Riggs, A. F. Bangura, C. Kucharczyk,
  M. Johannes, and I. R. Fisher, Phys. Rev. B {\bf 80}, 064507 (2009).

\bibitem{Yi2009b}M. Yi, D. H. Lu, J. G. Analytis, J.-H. Chu, S.-K. Mo,
  R.-H. He, M. Hashimoto, R. G. Moore, I. I. Mazin, D. J. Singh, Z. Hussain,
  I. R. Fisher, and Z.-X. Shen, Phys. Rev. B {\bf 80}, 174510 (2009).

\bibitem{Shimojima2010}T. Shimojima, K. Ishizaka, Y. Ishida, N. Katayama,
  K. Ohgushi, T. Kiss, M. Okawa, T. Togashi, X.-Y. Wang, C.-T. Chen,
  S. Watanabe, R. Kadota, T. Oguchi, A. Chainani, and S. Shin ,
  Phys. Rev. Lett. {\bf 104}, 057002 (2010).

\bibitem{McGuire2008}M. A. McGuire, A. D. Christianson, A. S. Sefat,
  B. C. Sales, M. D. Lumsden, R. Jin, E. A. Payzant, D. Mandrus, Y. Luan,
  V. Keppens, V. Varadarajan, J. W. Brill, R. P. Hermann, M. T. Sougrati,
  F. Grandjean, G. J. Long, Phys. Rev. B {\bf 78}, 094517 (2008);
  M. A. McGuire, R. P. Hermann, A. S. Sefat, B. C. Sales, R. Jin, D. Mandrus,
  F. Grandjean, and G. J. Long, New J. Phys. {\bf 11}, 025011 (2009).

\bibitem{Liu2008}R. H. Liu, G. Wu, T. Wu, D. F. Fang, H. Chen, S. Y. Li,
  K. Liu, Y. L. Xie, X. F. Wang, R. L. Yang, L. Ding, C. He, D. L. Feng, and
  X. H. Chen, Phys. Rev. Lett. {\bf 101}, 087001 (2008).

\bibitem{Dong2008}J. K. Dong, L. Ding, H. Wang, X. F. Wang, T. Wu, G. Wu,
X. H. Chen, and S. Y. Li, New J. Phys.\ \textbf{10}, 123031 (2008).

\bibitem{Drechsler2009}S.-L. Drechsler, H. Rosner, M. Grobosch, G. Behr,
  F. Roth, G. Fuchs, K. Koepernik, R. Schuster, J. Malek, S. Elgazzar,
  M. Rotter, D. Johrendt, H-H. Klauss, B. B\"{u}chner, and M. Knupfer,
  arXiv:0904.0827. 

\bibitem{WLYang2009}W. L. Yang, P. O. Velasco, J. D. Denlinger, A. P. Sorini,
  C-C. Chen, B. Moritz, W.-S. Lee, F. Vernay, B. Delley, J.-H. Chu,
  J. G. Analytis, I. R. Fisher, Z. A. Ren, J. Yang, W. Lu, Z. X. Zhao, J. van
  den Brink, Z. Hussain, Z.-X. Shen, and T. P. Devereaux, Phys. Rev. B {\bf
    80}, 014508 (2009).

\bibitem{chalcogenides}Despite their similar crystaline and electronic
  structure, it is likely that the iron chalcogenides are much more strongly
  correlated and have been suggested as better described by a local moment
  picture.~\cite{Johnston2010,Ma2009} 

\bibitem{Ma2009}F. Ma, J. Wei, J. Hu, Z.-Y. Lu, and T. Xiang, Phys. Rev. Lett.
  {\bf 102}, 177003 (2009).

\bibitem{Yildirim2008}T. Yildirim, Phys. Rev. Lett. {\bf 101}, 057010 (2008).

\bibitem{Uhrig2009}G. S. Uhrig, M. Holt, J. Oitmaa, O. Sushkov, and
  R. P. P. Singh, Phys. Rev. B {\bf 79}, 092416 (2009).

\bibitem{Krueger2009}F. Kr\"{u}ger, S. Kumar, J. Zaanen, and J. van den Brink,
  Phys. Rev. B {\bf 79}, 054504 (2009).

\bibitem{Schmidt2010}B. Schmidt, M. Siahatgar, and P. Thalmeier,
  Phys. Rev. B {\bf 81}, 165101 (2010).

\bibitem{Zhao2009}J. Zhao, D. T. Adroja, D.-X. Yao, R. Bewley, S. Li,
  X. F. Wang, G. Wu, X. H. Chen, J. Hu, and P. Dai, Nat. Phys. {\bf 5}, 555
  (2009). 

\bibitem{nesting}D. J. Singh and M.-H. Du, Phys. Rev. Lett. {\bf 100}, 237003
  (2008); I. I. Mazin, D. J. Singh, M. D. Johannes, and M. H. Du,
  \emph{ibid} {\bf 101}, 057003 (2008).

\bibitem{Zhang2010}Y.-Z. Zhang, I. Opahle, H. O. Jeschke, and R. Valent\'{\i},
  Phys. Rev. B {\bf 81}, 094505 (2010).

\bibitem{Korshunov2008}M. M. Korshunov and I. Eremin, Europhys. Lett. {\bf
  83}, 67003 (2008); Phys. Rev. B {\bf 78}, 140509(R) (2008).

\bibitem{Chubukov2008}A. V. Chubukov, D. Efremov, and I. Eremin, Phys. Rev. B
  {\bf{78}}, 134512 (2008).

\bibitem{Han2008}Q. Han, Y. Chen, and Z. D. Wang, Europhys. Lett. {\bf{82}},
  37007 (2008).

\bibitem{Vorontsov2009}A. B. Vorontsov, M. G. Vavilov, and A. V. Chubukov,
  Phys. Rev. B {\bf 79}, 060508(R) (2009).

\bibitem{Brydon2009a}P. M. R. Brydon and C. Timm, Phys. Rev. B {\bf 79},
  180504(R) (2009).

\bibitem{Cvetkovic2009}V. Cvetkovic and Z. Tesanovic Europhys. Lett. {\bf 85},
  37002, (2009); Phys. Rev. B {\bf 80}, 024512 (2009). 

\bibitem{Brydon2009b}P. M. R. Brydon and C. Timm, Phys. Rev. B {\bf 80},
  174401 (2009).

\bibitem{Thomale2009}R. Thomale, C. Platt, J. Hu, C. Honerkamp, and
  B. A. Bernevig, Phys. Rev. B {\bf 80}, 180505(R) (2009).

\bibitem{Knolle2010a}J. Knolle, I. Eremin, A. V. Chubukov, and R. Moessner,
  Phys. Rev. B {\bf 81}, 140506(R) (2010). 

\bibitem{Eremin2010}I. Eremin and A. V. Chubukov, Phys. Rev. B {\bf 81},
  024511 (2010).

\bibitem{Knolle2010b}J. Knolle, I. Eremin, A. Akbari, and R. Moessner,
  Phys. Rev. Lett. {\bf 104}, 257001 (2010). 

\bibitem{Fernandes2010}R. M. Fernandes and J. Schmalian, Phys. Rev. B {\bf
  82}, 014521 (2010).

\bibitem{Kuroki2008}K. Kuroki, S. Onari, R. Arita, H. Usui, Y. Tanaka,
  H. Kontani, and H. Aoki, Phys. Rev. Lett. {\bf 101},
  087004 (2008).

\bibitem{Lorenzana2008}J. Lorenzana, G. Seibold, C. Ortix, and M. Grilli,
  Phys. Rev. Lett. {\bf 101}, 186402 (2008).

\bibitem{Daghofer2008}M. Daghofer, A. Moreo, J. A. Riera, E. Arrigoni,
  D. J. Scalapino, and E. Dagotto, Phys. Rev. Lett. {\bf 101}, 
  237004 (2008).

\bibitem{Raghu2008}S. Raghu, Z.-L. Qi, C.-X. Liu, D. J. Scalapino, and
  S.-C. Zhang, Phys. Rev. B {\bf 77}, 220503(R) (2008).

\bibitem{Lee2008}P. A. Lee and X.-G. Wen, Phys. Rev. B {\bf 78}, 144517
  (2008).

\bibitem{Yanagi2008}Y. Yanagi, Y. Yamakawa, and Y. \=Ono,
  J. Phys. Soc. Jpn. {\bf 77}, 123701 (2008).

\bibitem{Ikeda2008}H. Ikeda, J. Phys. Soc. Jpn. {\bf 77}, 123707 (2008).

\bibitem{Kaneshita2009}E. Kaneshita, T. Morinari, and T. Tohyama,
  Phys. Rev. Lett. {\bf 103}, 247202 (2009).

\bibitem{Graser2009}S. Graser, T. A. Maier, P. J. Hirschfeld, and
  D. J. Scalapino, New J. Phys. {\bf 11}, 025016 (2009).

\bibitem{Kariyado2009}T. Kariyado and M. Ogata, J. Phys. Soc. Jpn. {\bf 78},
  043708 (2009).

\bibitem{Kubo2009}K. Kubo and P. Thalmeier, J. Phys. Soc. Jpn. {\bf 78},
  083704 (2009).

\bibitem{Arita2009}R. Arita and H. Ikeda, J. Phys. Soc. Jpn. {\bf 78}, 113707
  (2009). 

\bibitem{Ran2009}Y. Ran, F. Wang, H. Zhai, A. Vishwanath, and D.-H. Lee,
  Phys. Rev. B {\bf 79}, 014505 (2009). 

\bibitem{SLYu2009}S.-L. Yu, J. Kang, and J.-X. Li, Phys. Rev. B {\bf 79},
  064517 (2009).

\bibitem{Yu2009}R. Yu, K. T. Trinh, A. Moreo, M. Daghofer, J. A. Riera,
  S. Haas, and E. Dagotto, Phys. Rev. B {\bf 79}, 104510 (2009).

\bibitem{Moreo2009}A. Moreo, M. Daghofer, J. A. Riera, and E. Dagotto,
  Phys. Rev. B {\bf 79}, 134502 (2009). 

\bibitem{Kuroki2009}K. Kuroki, H. Usui, S. Onari, R. Arita, and H. Aoki,
  Phys. Rev. B {\bf 79}, 224511 (2009). 

\bibitem{Kaneshita2010}E. Kaneshita and T. Tohyama, Phys. Rev. B {\bf 82},
  094441 (2010).

\bibitem{Bascones2010}E. Bascones, M. J. Calder\'{o}n, and B. Valenzuela,
  Phys. Rev. Lett. {\bf 104}, 227201 (2010).

\bibitem{Daghofer2010a}M. Daghofer, A. Nicholson, A. Moreo, and E. Dagotto,
  Phys. Rev. B {\bf 81}, 014511 (2010). 

\bibitem{Ikeda2010}H. Ikeda, R. Arita, and J. Kune\v{s}, Phys. Rev. B {\bf
  81}, 054502 (2010).

\bibitem{Daghofer2010b}M. Daghofer, Q.-L. Luo, R. Yu, D. X. Yao, A. Moreo, and
  E. Dagotto, Phys. Rev. B {\bf 81}, 180514(R) (2010).

\bibitem{Long2010}M. S. Long, L. B. Hu, and W. LiMing, Eur. Phys. J. B {\bf
  75}, 205 (2010).

\bibitem{Thomale2010}R. Thomale, C. Platt, W. Hanke, and B. A. Bernevig,
  arXiv:1002.3599.

\bibitem{Luo2010}Q. Luo, G. Martins, D.-X.Yao, M. Daghofer, R. Yu, A. Moreo,
  and E. Dagotto, Phys. Rev. B {\bf 82}, 104508 (2010).

\bibitem{Zhou2010}S. Zhou and Z. Wang, Phys. Rev. Lett. {\bf 105}, 096401
  (2010). 

\bibitem{Yang2011}F. Yang, H. Zhai, F. Wang, and D.-H. Lee, Phys. Rev. B {\bf
  83}, 134502 (2010).

\bibitem{Kubo2010}K. Kubo and P. Thalmeier, arXiv:1010.4626.

\bibitem{Schickling2010}T. Schickling, F. Gebhard, and J. B\"{u}nemann,
  arXiv:1011.6219.

\bibitem{Excitonic}L. V. Keldysh and Y. V. Kopaev, Sov. Phys. Solid State {\bf
  6}, 2219 (1965); J. des Cloizeaux, J. Phys. Chem. Solids {\bf 26}, 259
  (1965); D. J\'erome, T. M. Rice, and W. Kohn, Phys. Rev. {\bf 158}, 462
  (1967).

\bibitem{Buker1981}D. W. Buker, Phys. Rev. B {\bf{24}}, 5713 (1981).

\bibitem{Boeri2008}L. Boeri, O. V. Dolgov, and A. A. Golubov,
  Phys. Rev. Lett. {\bf 101}, 026403 (2008).

\bibitem{Eschrig2009}H. Eschrig and K. Koepernik, Phys. Rev. B {\bf 80},
  104503 (2009).

\bibitem{ARPESZhang2009}Y. Zhang, B. Zhou, F. Chen, J. Wei, M. Xu, L. X. Yang,
  C. Fang, W. F. Tsai, G. H. Cao, Z. A. Xu, M. Arita, H. Hayashi, J. Jiang,
  H. Iwasawa, C.H. Hong, K. Shimada, H. Namatame, M. Taniguchi, J. P. Hu,
  and D. L. Feng, arXiv:0904.4022v2.

\bibitem{Ishida2009}K. Ishida, Y. Nakai, and H. Hosono,
  J. Phys. Soc. Jpn. {\bf 78}, 062001 (2009).

\bibitem{Si2009}Q. Si, E. Abrahams, J. Dai, and J.-X. Zhu, New. J. Phys. {\bf
  11}, 045001 (2009).

\bibitem{Aichhorn2009}M. Aichhorn, L. Pourovskii, V. Vildosola, M. Ferrero,
  O. Parcollet, T. Miyake, A. Georges, and S. Biermann, Phys. Rev. B {\bf 80},
  085101 (2009).

\bibitem{Calderon2009a}M. J. Calder\'{o}n, B. Valenzuela, and E. Bascones,
  New J. Phys. {\bf 11}, 013051 (2009).

\bibitem{Calderon2009b}M. J. Calder\'{o}n, B. Valenzuela, and E. Bascones,
  Phys. Rev. B {\bf 80}, 094531 (2009).

\bibitem{Brydon2011}P. M. R. Brydon, M. Daghofer, C. Timm, and J. van den
  Brink, Phys. Rev. B {\bf 83}, 060501(R) (2011).

\bibitem{Manousakis2008}E. Manousakis, J. Ren, S. Meng, and E. Kaxiras,
  Phys. Rev. B {\bf 78}, 205112 (2008).

\bibitem{Papaconstantopoulos2010}D. A. Papaconstantopoulos, M. J. Mehl, and
  M. D. Johannes, Phys. Rev. B {\bf 82}, 054503 (2010).

\bibitem{Oles1983}A. M. Ole{\'s}, Phys. Rev. B {\bf 28}, 327 (1983).

\bibitem{Nomura2000}T. Nomura and K. Yamada, J. Phys. Soc. Jpn. {\bf 69}, 1856
  (2000).  

\bibitem{2bandnote}The tight-binding band structure of the two-orbital model
  is parameterized in terms of the nearest-neighbour hopping integral
  $t_1$,~\cite{Raghu2008} which we here take as $t_1=1$\,eV. Although
  a smaller value of $t_1$ would give a more realistic bandwidth, this only
  results in a rescaling of the $T$-$U$ phase diagram and does not alter the
  underlying physics. 

\bibitem{LiFeAsARPES}S. V. Borisenko, V. B. Zabolotnyy, D. V. Evtushinsky,
  T. K. Kim, I. V. Morozov, A. N. Yaresko, A. A. Kordyuk, G. Behr,
  A. Vasiliev, R. Follath, and B. B\"{u}chner, Phys. Rev. Lett. {\bf 105},
  067002 (2010). 

\bibitem{LiFeAsNMR}S. H. Baek, H. J. Grafe, F. Hammerath, M. Fuchs, L. Harnagea,
  S. Wurmehl, J. van den Brink, and B. B\"{u}chner (unpublished).

\bibitem{Grasernote}Note that the this hole pocket is not visible
  in~\Ref{Graser2009}, as the ``undoped'' compound in this paper actually
  corresponds to a slight electron doping, $n=6.032$.

\bibitem{Yi2009a}M. Yi, D. H. Lu, J. G. Analytis, J.-H. Chu,
  S.-K. Mo, R.-H. He, R. G. Moore, X. J. Zhou, G. F. Chen,
  J. L. Luo, N. L. Wang, Z. Hussain, D. J. Singh, I. R. Fisher, and
  Z.-X. Shen, Phys. Rev. B {\bf 80}, 024515 (2009).

\bibitem{Yildirim2009}T. Yildirim, Phys. Rev. Lett. {\bf 102}, 037003
  (2009). 

\bibitem{Mazin2008}I.I. Mazin, M.D. Johannes, L. Boeri, K. Koepernik, and
  D.J. Singh, Phys. Rev. B {\bf 78}, 085104 (2008).

\bibitem{Vildosola2008}V. Vildosola, L. Pourovskii, R. Arita, S. Biermann, and
  A. Georges, Phys. Rev. B {\bf 78}, 064518 (2008).

\bibitem{Cao2008}C. Cao, P. J. Hirschfeld, and H.-P. Cheng, Phys. Rev. B {\bf
  77}, 220506(R) (2008).

\bibitem{Barzykin2009}V. Barzykin and L. P. Gor'kov, Phys. Rev. B {\bf 79},
  134510 (2009).

\bibitem{Lv2009}W. Lv, J. Wu, and P. Phillips, Phys. Rev. B {\bf 80}, 224506
  (2009). 

\bibitem{Chen2010}C.-C. Chen, J. Maciejko, A. P. Sorini, B. Moritz,
  R. R. P. Singh, and T. P. Devereaux, Phys. Rev. B {\bf 82}, 100504(R) (2010).

\bibitem{Hansmann2010}P. Hansmann, R. Arita, A. Toschi, S. Sakai,
  G. Sangiovanni, and K. Held, Phys. Rev. Lett. {\bf 104}, 197002 (2010).



\end{thebibliography}
\end{document}